\documentclass[twocolumn,showkeys,prd,nofootinbib,floatfix,preprintnumbers]{revtex4-1}

\usepackage[utf8]{inputenc}
\usepackage{multirow}
\usepackage{amsfonts,amsmath,amssymb} 
\usepackage{graphicx,graphics,color}
\usepackage{gensymb}
\usepackage{longtable}
\usepackage{bbding}
\usepackage{subfigure}

\usepackage{hyperref}
\usepackage[normalem]{ulem}
\hypersetup{
    colorlinks=true,
    linkcolor=blue}

\def\apss{Ap\&SS}   
  
  \def\prd{Phys. Rev. D}

\begin{document}

\title{Accretion of self-interacting scalar field dark matter onto a Reissner-Nordstr\"{o}m black hole}

\author{Yuri Ravanal}
\email{yuri.ravanal@usach.cl}
\affiliation{Departamento de F\'isica, Universidad de Santiago de Chile,\\Avenida V\'ictor Jara 3493, Estaci\'on Central, 9170124 Santiago, Chile}

\author{Gabriel G\'omez}
\email{gabriel.gomez.d@usach.cl}
\affiliation{Departamento de F\'isica, Universidad de Santiago de Chile,\\Avenida V\'ictor Jara 3493, Estaci\'on Central, 9170124 Santiago, Chile}

\author{Normal Cruz}
\email{norman.cruz@usach.cl}
\affiliation{Departamento de F\'isica, Universidad de Santiago de Chile,\\Avenida V\'ictor Jara 3493, Estaci\'on Central, 9170124 Santiago, Chile}
\affiliation{Center for Interdisciplinary Research in Astrophysics and Space Exploration (CIRAS), Universidad de Santiago de Chile, Avenida Libertador Bernardo O’Higgins 3363, Estación Central, Chile}

\begin{abstract}
Self-interacting scalar field dark matter can be seen as an extension of the free case known as fuzzy dark matter. The interactive case is capable of reproducing the positive features of the free case at both astrophysical and cosmological scales. 
On the other hand, current imaging black holes (BHs) observations provided by the Event Horizon Telescope (EHT) collaboration cannot rule out the possibility that BHs can carry some amount of charge. Motivated by these aspects, and by the possibility of detecting dark matter through its gravitational imprints on BH observations, in this paper, we extend previous studies of accretion of self-interacting scalar field dark matter to the charged BH case. Our analysis is based on the assumption on spherically symmetric flow and employs a test fluid approximation. All analytical expressions are derived from the ground up in Schwarzschild coordinates. Concretely, we implement analytical and numerical approaches to investigate the impact of the charge on the energy flux. From this analysis, we notice that the mass accretion  rate efficiency is reduced up to $\sim 20\%$ for the maximum allowed charge. Additionally, considering the mass accretion rate of M87$^{\star}$ inferred from polarization data of the EHT, we infer the conservative bound 
$ \lambda_4 > (1.49-10.2)( m / 1 \rm  {eV} )^4$ based on the simple criterion that ensures the mass accretion rate caused by DM remains subdominant compared to the baryonic component.

\end{abstract}

\maketitle

\section{Introduction}

Observational studies of the cosmic microwave background (CMB) anisotropies \cite{WMAP:2003elm,Planck:2015fie,Planck:2018vyg}, and the formation of large-scale structures (LSS) \cite{SDSS:2003tbn,BOSS:2013rlg,BOSS:2016wmc}, indicate that a non-luminous component constitutes approximately $85\%$ of matter in the Universe. The standard cosmology model $\Lambda$CDM describes the dark sector using two main components: dark energy ($\Lambda$), responsible for the accelerated expansion of the Universe; and cold dark matter (CDM), responsible for the formation of structures at different scales. The CDM model describes the DM component as a nonrelativistic perfect fluid. While the $\Lambda$CDM model has been very successful in describing large-scale observations, $N$-body (DM only) simulations have difficulties in describing structures at (sub) galactic scales, resulting in the ``core-cusp" \cite{Moore:1994yx,deBlok:2009sp}, the ``missing satellites" \cite{Moore:1999nt,Bullock:2000wn} and the ``too big to fail" problems \cite{Boylan-Kolchin:2011qkt,Garrison-Kimmel:2014vqa}. Some investigations, however, have indicated that such issues could be alleviated if simulations take into account the physics associated with baryons \cite{DelPopolo:2016emo,Dutton:2018nop,Dutton:2020vne}. Alternatively, exploring alternative models for DM is a promising approach to address these issues and it is also motivated by the lack of a robust observational signature in current experimental facilities.

The Scalar-Field Dark Matter (SFDM) model, which proposes that DM is composed of ultra-light bosonic (spin-0) particles with a mass $ m \thicksim [10^{-22}-1]$ eV, has gained significant attention in recent times due to the number of problems it might help to solve \cite{Hu:2000ke,Hui:2016ltb,Ferreira:2020fam}. The growing interest of this model stems from the fact that these particles possess de Broglie wavelengths that are comparable to astrophysical scales. 

There are various categories within the SFDM model, such as Axions \cite{Peccei:1977hh,Wilczek:1977pj,Weinberg:1977ma}, Fuzzy Dark Matter (FDM) \cite{Hu:2000ke,Hui:2016ltb,Sin:1992bg,Ji:1994xh,Seidel:1990jh,Matos:1998vk,Hui:2019aqm}, and Self-Interacting Scalar Field Dark Matter (SIDM) \cite{Lee:1995af,Goodman:2000tg,Peebles:2000yy,Arbey:2003sj,Boehmer:2007um,Lee:2008jp,Harko:2011xw,Rindler-Daller:2011afd,Suarez:2013iw,Urena-Lopez:2019kud}. The primary difference between these models lies in their associated masses and couplings ranges \cite{Kobayashi:2017jcf,Abel:2017rtm,Brito:2017wnc,Brito:2017zvb}. An attractive feature of SFDM models is that they can form soliton cores at galactic centers, offering a plausible explanation for the observed DM cores in (sub) halos 
\cite{Arbey:2001qi,Schive:2014hza,Marsh:2015wka,Schwabe:2016rze,Mocz:2017wlg,Levkov:2018kau,2011PhRvD..84d3531C,2011PhRvD..84d3532C}. The soliton cores can have radii between $[1-20]$ kpc, depending on the particle mass. In addition to explaining DM cores, SFDM models can also account for the origin of vortices in galaxies \cite{Rindler-Daller:2011afd,Peebles:1969jm,Kain:2010rb}. However, recent reports suggest that the FDM model may face challenges when dealing with the Lyman-$\alpha$ forest data \cite{Kobayashi:2017jcf,Irsic:2017yje,Rogers:2020ltq} and rotation curves data from SPARC database \cite{Bar:2021kti} for particle masses around $m\sim 10^{-22}$ eV. Although it has been recently argued that the inclusion of small but a nonvanishing self-interaction may help to reconcile predictions with rotation curves data \cite{Dave:2023wjq}.

The presence of supermassive black holes (SMBHs) in most galaxies is well-established and supported by observations \cite{Kormendy:1995er,Ferrarese:2004qr,Narayan:2005ie}. Moreover, certain constraints suggest the possibility that BHs can carry some amount of charge \cite{Zakharov:2014lqa,Juraeva:2021gwb}. Recent results obtained in the strong field regime by the Event Horizon Telescope (EHT) collaboration \cite{EventHorizonTelescope:2021dqv,EventHorizonTelescope:2022xqj}, which captured for the first time the shadow image of M87$^{\star}$ \cite{EventHorizonTelescope:2021dqv}, have further confirmed this possibility. Subsequently, data obtained by the EHT for Sagittarius A$^{\star}$ (SgrA$^{\star}$) \cite{EventHorizonTelescope:2022xqj,Vagnozzi:2022moj} further support this idea.  From these observations, a charge-mass ratio $q\equiv Q/M$ was obtained for M87$^{\star}$ to be in the range $q \in [0, 0.9]$ \cite{EventHorizonTelescope:2021dqv}, and $q \in [0, 0.84]$ for SgrA$^{\star}$ \cite{EventHorizonTelescope:2022xqj}. However, it is widely believed that astrophysical BHs are electrically neutral due to charge neutralization by astrophysical plasma. It is important to mention that this article does not discuss the origin of the BH charge, and only accepts the aforementioned constraints as a probe of concept.

One promising strategy to detect any potential signature of DM is to search for its gravitational effects on BH observations (see e.g \cite{Eda:2013gg,Lacroix:2012nz,Kavanagh:2020cfn,Gomez:2016iod,Bar:2019pnz,Saurabh:2020zqg,Boudon:2023vzl,Chakrabarti:2022owq}). However, the effect of DM on BH observations strongly depends on the DM distribution, which provides a promising opportunity to gain insights into the nature of DM \cite{Macedo:2013qea,Barausse:2014tra}. Therefore, it is crucial to perform an accurate and consistent modeling of DM around BHs (see e.g. \cite{Sadeghian:2013laa} for the CDM case). In the context of SFDM scenarios, some studies have investigated the gravitational influence of BHs on the scalar field profile \cite{Barranco:2011eyw,Clough:2019jpm,Bamber:2020bpu,Hui:2019aqm,Aguilar-Nieto:2022jio,Brax:2019npi,Vicente:2022ivh,Cardoso:2022nzc}. In particular, Ref.~\cite{Brax:2019npi} determined the scalar field profile for the SIDM model by demanding steady energy flux onto a central Schwarzschild BH in the strong field regime and considering the mass large limit,  where $ m \gg 10^{-22}$eV \cite{Brax:2019fzb}. In this limit, the quantum pressure can be neglected, meaning that the repulsive self-interaction plays a crucial role in maintaining the equilibrium of the scalar cloud on subgalactic scales\footnote{The term scalar cloud is employed here to describe a finite-sized macroscopic structure surrounding a BH. This long-lived self-gravity structure is stable due to the repulsive self-interaction.}  \cite{Chavanis:2018pkx,Khlopov:1985jw,Goodman:2000tg,Li:2013nal,Suarez:2015fga,Suarez:2016eez,Suarez:2017mav}. These are some of the theoretical considerations that we keep in mind in the present paper. 

In this paper, we aim to extend the results obtained in Ref.~\cite{Brax:2019npi}, considering the effects of BH charge on the energy flux. Specifically, we investigate the accretion of self-interacting scalar field DM around a Reissner-Nordstr\"{o}m (RN) BH \cite{reissner1916eigengravitation}. To allow further investigation of our results in other astrophysical scenarios, we derive all analytical expressions from the ground up in Schwarzschild coordinates. Our proposal is mainly motivated by observations of the EHT, which allows for the existence of a more general class of BHs, including those endowed with electric charge. Thus, we focus on the interplay between the BH charge and the self-interacting scalar field which could have implications for the behavior of DM around BHs in a general astrophysical scenario.

We found that different critical flow rates occur for various values of the BH charge. As the charge increases, the critical flux decreases. However, there is a unique transonic flow that is independent of the charge, indicating that the flow moves on a single critical curve determined by the stability criteria. Furthermore, considering that $\dot M_{\rm baryons} \gg \dot M_{\rm DM}$, we obtained a conservative constraint on the parameter space ($m$, $\lambda_4$) by using the observed accretion mass rate of M87$^{\star}$. Although we observed a noticeable difference in the accretion process when the charge becomes important, it remains subdominant compared to what is observed in baryon accretion \cite{Salpeter:1964kb}. It is important to note that we only considered radial accretion in this article and did not include the backreaction effect of baryons.

This paper is structured as follows. In Sec. \ref{sec:II}, we describe the theoretical framework, including the DM model, spacetime geometry, and equations of motion. Section \ref{sec:III} presents the numerical results exploring the impact of the BH charge on the critical flow, density profile, and DM accretion rate. Finally, in Sec. \ref{sec:IV}, we provide a detailed discussion of our results and their potential implications.

\section{Dark Matter Scalar Field}
\label{sec:II}

The scalar-field action with minimal coupling to gravity is given by

\begin{equation}
    S_\phi = \int d^4x \sqrt{-g} \left[ - \frac{1}{2} g^{\mu\nu} \partial_\mu\phi \partial_\nu\phi
- V(\phi) \right].\label{action}
\end{equation}

The first term inside the bracket represents the kinetic term and the second term represents the potential. In this study, we use the metric signature conventions $(-,+,+,+)$ and units $ 4\pi\epsilon_{0} = G = c = \hbar = 1$.

The equation of motion for the SF, derived from Eq. (\ref{action}), takes the following form
\begin{equation}
\frac{\delta S_\phi }{\delta \phi} = 0 \;\;\; \Longrightarrow \;\;\; \Box \phi - \frac{dV}{d\phi} = 0,\label{Equation motion}
\end{equation}
where $\Box=\nabla^{\mu}\nabla_{\mu}=g^{\mu\nu}\nabla_{\mu}\nabla_{\nu}$ is the covariant d’Alembertian, $ V(\phi)=m^2 \phi^2/2 + V_I(\phi) $ is the scalar field potential, and $V_I(\phi)=\lambda_4\phi^4/4$ is a repulsive quartic self-interaction term. In Minkowski spacetime and in the absence of self-interaction, the usual Klein-Gordon equation $(\Box-m^2)\phi=0$ is recovered. In this paper, we consider a range of masses in the interval $ 10^{-19}\;\rm{eV} \ll m \thicksim 1 $ eV \cite{Brax:2019npi}, which corresponds to the regime of mass-large, where the quantum pressure contributions at both galactic and subgalactic levels can be safely neglected. In such a scenario, solitonic nuclei formed in galactic centers may have originated from the SIDM scenario, where the scalar cloud collapses due to purely gravitational effects, and the mechanism that stops this collapse is the repulsive self-interaction, leading to an equilibrium state.

Before deriving the master equations, it is important to mention the main physical assumptions considered in this work for the sake of clarity:
\begin{itemize}
    \item Radial accretion flows on static spherically symmetric BHs.
    \item large scalar mass limit where the Compton wavelength $1/m$ is smaller than the BH size.
    \item Test-fluid approximation where the backreaction of the scalar cloud to the spacetime metric is neglected. 
\end{itemize}

\subsection{Spherically
symmetric space-times}

For a spherically symmetric spacetime the metric takes the form
 \begin{equation}
 ds^2 = - f(r) dt^2 + g(r) dr^2 + r^2  d\vec\Omega^2,\label{metric SS}
 \end{equation}
 where the metric functions $f(r)$ and $g(r)$ for a RN metric are given by
 \begin{equation}
 f(r)=\frac{1}{g(r)}=\left(1-\frac{2M}{r}+\frac{Q^2}{r^2}\right),\label{coef}
 \end{equation}
with the corresponding horizons
 \begin{equation}
 r_{\pm} = M \pm \sqrt{M^2-Q^2}.\label{horizont}
 \end{equation}
Here $M$ is the BH mass and $Q$ is the associated electric charge. The BH metric displays two event horizons: $r_+$ corresponds to the external horizon (which is the one of interest in this work), and $r_-$ corresponds to the Cauchy horizon. The Schwarzschild solution is straightforwardly recovered when Q = 0, and the case $Q=M$ describes the extremal case. The metric functions can be rewritten trivially using the following change of variables
\begin{equation}
x=\frac{r}{M} \geq 1,\; \text{and} \;\;\;q=\frac{Q}{M}.\label{Change Variables}
\end{equation}
Here, $x$ represents our new dimensionless radial coordinate, and $q$ represents the (dimensionless) charge-to-mass ratio. Using these variables, we can express the horizons as follows
\begin{equation}
    x_{\pm} = 1 \pm \sqrt{1-q^2} : \;\;\; 0\leq q \leq 1.\label{new horizont}
\end{equation}
As discussed in Ref.~\cite{Brax:2019npi}, there are three distinct regions of interest. The first region refers to the strong-gravity regime in the vicinity of the BH, where the metric functions have a dominant influence. This region extends from the horizon $r_+$ up to a radius $r_{\rm NL}$. The second region corresponds to the weak-gravity regime, in which the effects of the BH are nearly Newtonian ($\Phi=-M/r\ll 1$) and extends from a very far distance from the BH up to a radius $r_{\rm sg}$. Finally, the third region is a region in which
self-gravity of the DM cloud dominates the gravitational potential in the Poisson equation
\begin{equation}
r \gg r_{\rm sg}  : \;\;\; \nabla^2 \Phi = 4\pi \rho_\phi.\label{ecuacion de poisson}
\end{equation}
Here $\Phi$ denotes the Newtonian gravitational potential, while $\rho_\phi$ stands for the energy density of the SF.

\subsection{Equations of motion}

In this section, we closely follow the calculations made by \cite{Brax:2019npi} and only show the main expressions that present changes as a result of the new coordinates system chosen. For the case of quartic self-interactions, we can use Eq. (\ref{Equation motion}) and the metric defined by Eqs. (\ref{metric SS}) and (\ref{coef}) to write the following nonlinear Klein-Gordon equation in Schwarzschild coordinates
\begin{equation}
\frac{\partial^2\phi}{\partial t^2} - \sqrt{\frac{f}{g}} \frac{1}{r^2} \frac{\partial}{\partial r}
\left[ \sqrt{\frac{f}{g}}r^2\frac{\partial\phi}{\partial r} \right] + f m^2 \phi + f \lambda_4 \phi^3 = 0.\label{K-G nolineal}
 \end{equation}
 This equation can be identified as a Duffing-type equation \cite{kovacic2011duffing} in the large scalar mass limit since the characteristic length scale of the system is larger than the Compton wavelength $\lambda_{C}\sim 1/m$. This equation describes a harmonic oscillator with nonlinear (cubic) restoring force, and no damping or driving. The exact solution to this equation is given by the Jacobi elliptic functions $y=Y ep (u,k)$ \cite{gradshteyn19651,byrd1971table}. Here $ep$ is a general expression describing the Jacobi elliptic sine ${(\rm sn)}$, cosine ${(\rm cn)}$, and delta ${(\rm dn)}$ functions with argument $u=\omega t-\beta$ and modulus $k$. In this context, $\beta$ represents the phase, and holds a direct relationship with the radial velocity (we will delve further into this later on). Conversely, the modulus $k$ denotes the nonlinear deviations of the harmonic oscillator, which characterizes the relativistic regime. These functions are doubly periodic, with a period of 4${\bf K}$, where ${\bf K}(k) = \int_0^{\pi/2} d\theta /\sqrt{1-k^2\sin^2\theta}$ and ${\bf E}(k) = \int_0^{\pi/2} d\theta \sqrt{1-k^2\sin^2\theta}$ are the complete elliptic integrals of the first and second kind, respectively, defined within the interval $k \in [ 0, 1 ) $. Solution of Eq. (\ref{K-G nolineal}) can be written as follows \cite{Brax:2019npi,Frasca:2009bc}
 
 \begin{equation}
     \phi(r,t) = \phi_0(r) \, {\rm cn}[ \omega(r) t - {\bf K}(r) \beta(r), k(r) ].\label{Solucion K-G no lineal}
 \end{equation}
 
 In the limit of large scalar mass, the radial derivatives of both the amplitude $\phi_0$ and the modulus $k$ are significantly smaller than $ m $ ($ \partial_r \ll m $). Meanwhile, the angular frequency $\omega$ and the phase $\beta$ have the same order of magnitude as $m$. Additionally, it is required that the field oscillates in phase with a period $T=2\pi/\omega_0$, where $\omega_0$ is the common angular frequency and $\omega (r) = 4{\bf K}(r)/T$. If the field does not oscillate in phase, a growth may occur, leading to an increase in the radial derivatives. In this limit, the Jacobi elliptic function ${(\rm cn)}$ can be expressed as a Fourier expansion \cite{byrd1971table}. Therefore, the temporal and radial derivatives of $\phi(r,t)$ can be related to the derivatives of the Jacobi elliptic functions in the following way
 \begin{equation}
 \frac{\partial\phi}{\partial t} = \phi_0 \omega \frac{\partial {\rm cn}}{\partial u},\label{relacion temporal phi}
 \end{equation}
 \begin{equation}
 \frac{\partial\phi}{\partial r} = - \phi_0 {\bf K} \beta' \frac{\partial {\rm cn}}{\partial u} + \dots,\label{relacion radial phi}
 \end{equation}
where dots represent subdominant terms and $\beta' = d\beta/dr$. By substituting Eqs. (\ref{relacion temporal phi}) and (\ref{relacion radial phi}) into Eq.(\ref{K-G nolineal}), we obtain the following result
 
\begin{equation}
    \phi_0 \left[ \omega^2 - \frac{f}{g} ( {\bf K} \beta' )^2 \right] \frac{\partial^2 {\rm cn}}{\partial u^2}
+ f m^2 \phi_0 {\rm cn} + f \lambda_4 \phi_0^3 {\rm cn}^3 = 0,\label{K-G jacobi}
\end{equation}
which exhibits a similar structure  of the Duffing's equation. Employing the property ${\partial^2 {\rm cn}}/{\partial u^2} = (2k^2-1) {\rm cn} - 2 k^2 {\rm cn}^3$, we can construct an algebraic equation for the factors ${\rm cn}$ and ${\rm cn}^3$, which can be expressed as follows

\begin{equation}
    \frac{\pi^2 f}{4 g} \beta'^2 = \omega_0^2 - \frac{f m^2 \pi^2}{(1-2k^2) 4 {\bf K}^2}.\label{G Euler equation}
\end{equation}
\begin{equation}
    \frac{\lambda_4 \phi_0^2}{m^2} = \frac{2k^2}{1-2k^2}.\label{Relacion Duffing}
\end{equation}
Eqs. (\ref{G Euler equation}) and (\ref{Relacion Duffing}) provide two conditions for the self-interaction and mass of the SF. 

In the limit of vanishing self-interaction ($\lambda_4 \rightarrow 0$), we have $k \rightarrow 0$, which implies that the function ${\rm cn}(u,0)$ approaches ${\rm cos}(u)$. It is important to note that the term $\pi \beta'/(2m)$ represents the radial velocity $v_r$, which will be relevant for the accretion flow. Additionally, we observe a singularity at $k_\pm = \pm 1/\sqrt{2}$, where we exclude the branch $k_- = -1/\sqrt{2}$ since $k$ is defined on the interval $[0,1)$, and only $k_+$ is relevant. As a result, a divergence around $k_{+}$ is expected in the numerical calculation of the accretion flow. Moreover, for $k > 1/\sqrt{2}$, we observe a change of sign in Eqs. (\ref{G Euler equation}) and (\ref{Relacion Duffing}). This behavior is similar to the one observed when demanding regularity in the accretion flow of a polytropic fluid  \cite{Bondi:1952ni,1972Ap&SS..15..153M}. Therefore, this value sets the boundary at which transonic flow is allowed, and it has significant physical consequences for the study of stable accretion flow, as we will see later.

\subsubsection{Solitonic conditions, steady-state, and constant flux.}

A real SF $\phi$ can be expressed as the sum of two complex SF's $ \phi = (e^{-i m t}\psi + e^{i m t}\psi^*)/\sqrt{2m}$. Using the Madelung transformation \cite{madelung1927quantum}, we can bring this problem into the hydrodynamic framework and derive the continuity and Euler equations. For more details, we refer the reader to \cite{Brax:2019npi}. For large distances $r \gg r_{\rm sg} $ within the weak-gravity regime, the scalar cloud is dominated by self-gravity. Thus, demanding regular boundary conditions, we have to connect the solitonic solution with solution Eq. (\ref{Solucion K-G no lineal}). Moreover the soliton exhibits hydrostatic equilibrium ($\vec v \sim 0$), implying that $\vec\nabla (\Phi+\Phi_{\rm I}) =0$. Direct integration provides

\begin{equation}
    r \leq R_s : \;\;\; \Phi+\Phi_{\rm I} = \alpha,\label{definicion alfa}
\end{equation}
where $R_s$ represents the radius of the soliton, $\Phi_{\rm I}$ denotes the repulsive potential resulting from self-interaction, and $\alpha$ is an integration constant. According to \cite{Brax:2019fzb}, $\Phi_{\rm I}(\rho) = \rho / \rho_a$, where $ \rho_a = 4m^4/3\lambda_4 $. The solution of Eq (\ref{definicion alfa}) can be written as

\begin{equation}
    \psi = \sqrt{\frac{\rho}{m}} e^{-i\alpha m t}  \;\;\; \Longrightarrow \;\;\; s= - \alpha m t,
\end{equation}
and
\begin{equation}
    \phi = \frac{\sqrt{2\rho}}{m} \cos[ (1+\alpha) m t ].\label{solucion real soliton}
\end{equation}
Furthermore, within the soliton $r_{sg} \ll r \ll R_s$, the self-interaction potential $V_{\rm I} \sim \rho \Phi_{\rm I} \ll \rho$. In other words, $\lambda_4 \phi^4 \ll m^2 \phi^2 $. We can expand Eq. (\ref{Relacion Duffing}) for $ k \ll 1$ and obtain the following expression

\begin{equation}
    k^2 = \frac{\lambda_4 \phi_0^2}{2m^2} + \dots,\label{k2 muy chico}
\end{equation}
where dots represent next to leading contributions. Following this same idea and using the Fourier series expansion of the Jacobi elliptic function \cite{byrd1971table}, we obtain the following expression valid for $ k \ll 1$
\begin{equation}
    \phi = \phi_0 \cos( \omega_0 t - \pi \beta/2) + \dots.\label{rho solution k2 pequeño}
\end{equation}
We can compare Eqs. (\ref{solucion real soliton}) and (\ref{rho solution k2 pequeño}). Additionally , we have that $\beta$ is related to velocity through the Madelung transformations \cite{madelung1927quantum}, and for $\beta \sim$ 0, we find that $ \phi_0 = \sqrt{2\rho}/m $ and $\omega_0 = (1+\alpha)m $. This result is significant because it implies that $ \alpha = \Phi_{\rm I}(R_s) \sim 10^{-5} $ within the soliton, and typically $\Phi$ is in the range of [$10^{-6}$ - $10^{-5}$] for cosmological and galactic scales.

We can express the conservation of energy and momentum in the relativistic framework through the equation $\nabla_\mu T^\mu_\nu = 0$, where the time component ($\nu=t$) corresponds to the continuity equation. The nonvanishing components of the energy-momentum tensor for the SF are given by 
\begin{equation}
\rho_\phi \equiv - T^t_t = \frac{1}{2f} \left( \frac{\partial\phi}{\partial t} \right)^2
+ \frac{1}{2g} \left( \frac{\partial\phi}{\partial r} \right)^2 + V(\phi).\label{rho_scalar}
\end{equation}
and
\begin{equation}
T^r_t = \frac{1}{g} \frac{\partial\phi}{\partial r} \frac{\partial\phi}{\partial t}.\label{Componente T00}  
\end{equation}
By using Eqs. (\ref{Solucion K-G no lineal}), (\ref{relacion temporal phi}), (\ref{relacion radial phi}), (\ref{G Euler equation}) and (\ref{Relacion Duffing}), Eqs. (\ref{rho_scalar}) and (\ref{Componente T00}) can be written in terms of the Jacobi elliptic functions as
\begin{equation}
    \rho_\phi= \frac{(1-k^2)m^2\phi_0^2}{2(1-2k^2)} + \phi_0^2 \frac{({\bf K}\beta')^2}{g}
[ 1 - k^2 + (2k^2-1) {\rm cn}^2 - k^2 {\rm cn}^4 ],\label{rho algebraico}
\end{equation}
and
\begin{equation}
    T^r_t = - \phi_0^2 \omega \frac{{\bf K}\beta'}{g} \left( \frac{\partial {\rm cn}}{\partial u} \right)^2,\label{T00 jacobi}
\end{equation}
where $({\partial {\rm cn}}/{\partial u})^2= 1 - k^2 + (2k^2-1) {\rm cn}^2 - k^2 {\rm cn}^4$. The continuity equation can be expressed as follows
\begin{equation}
    \dot\rho - \frac{1}{\sqrt{f g} r^2} \frac{\partial}{\partial r} \left[ \sqrt{f g} r^2 T^r_t \right] = 0,\label{ecuacion de continudad T00}
\end{equation}
where dot denotes the time derivative. The Jacobi functions exhibit periodic behavior, therefore Eq. (\ref{rho algebraico}) is not constant over time. As we seek for steady state and constant flow conditions, it is convenient to average $\langle ... \rangle$ over one oscillation period $T = 2\pi/\omega_0$. Specifically, the averaging process involves the square $ \langle {\rm cn}^2 \rangle $ and fourth $ \langle {\rm cn}^4 \rangle $ powers of the Jacobi function. These averages are denoted, respectively, as $C_2$ and $C_4$, where $ C_2 = ({\bf E}/{\bf K} + k^2-1)/k^2 $ and $ C_4 = [2(2k^2-1)C_2 +1 - k^2]/3k^2 $. With all this in mind, we can calculate the energy flux resulting from the steady-state as follows

\begin{equation}
    F =\sqrt{\frac{f}{g}} r^2 \phi_0^2 \omega {\bf K} \beta'
\left \langle \left( \frac{\partial {\rm cn}}{\partial u} \right)^2 \right\rangle.\label{flujo 1}
\end{equation}
We can determine the radial velocity $v_r$ from the Euler equation Eq. (\ref{G Euler equation}) with $ \omega(r)= 2{\bf K}(r)\omega_0/\pi $ and $ \omega_0 = (1+\alpha)m $. It yields
\begin{equation}
    v_r = \pm (1+\alpha) \sqrt{ \frac{g}{f} }
\sqrt{ 1 - \frac{\pi^2 f}{(1-2k^2) 4 {\bf K}^2(1+\alpha)^2} }.\label{velocidad}
\end{equation}
The minus sign is relevant for describing the (radial) infall of the SF onto the RN BH. Using Eqs. (\ref{G Euler equation}) and (\ref{Relacion Duffing}) and applying the previous definitions of $\omega(r)$ and $\omega_0$, the energy flux can be expressed, after some algebraic manipulations, as
\begin{widetext}
\begin{equation}
    F=F_{RN} x^2
 \left( \frac{2 {\bf K}}{\pi} \right)^2
[1 - k^2 + (2k^2-1) C_2
- k^2 C_4]
\frac{k^2}{2(1-2 k^2)}\sqrt{ 1 - \frac{\pi^2 f}{(1-2k^2)4 {\bf K}^2 (1+\alpha)^2 } }.\label{Flujo final}
\end{equation}
\end{widetext}
Here, we have defined the characteristic flow for the RN case as 
\begin{equation}
  F_{RN} = - \frac{(2M)^2 m^4 (1+\alpha)^2}{\lambda_4} \thicksim - \frac{(2M)^2 m^4}{\lambda_4},\label{definicion F_RN}  
\end{equation}
which will be useful for normalization purposes in the subsequent numerical analysis. It is worth noting that Eq. (\ref{Flujo final}) has the usual form of a flow equation, where $F \sim \rho r^2 v_r$.  Lastly, it is important to emphasize that the energy flux is, in general, a function of three quantities: the modulus $k$, the radial coordinate $x$ and the (dimensionless) charge $q$, i.e., $F=F(k,x,q)$, which all together determine the behavior of the accretion flow. This is the main concern that we shall focus on in the next section.

\section{Numerical Results}\label{sec:III}

\subsection{Analysis of the function $F(k,x,q)$}

\begin{figure*}
    \centering
    \includegraphics[scale=0.4]{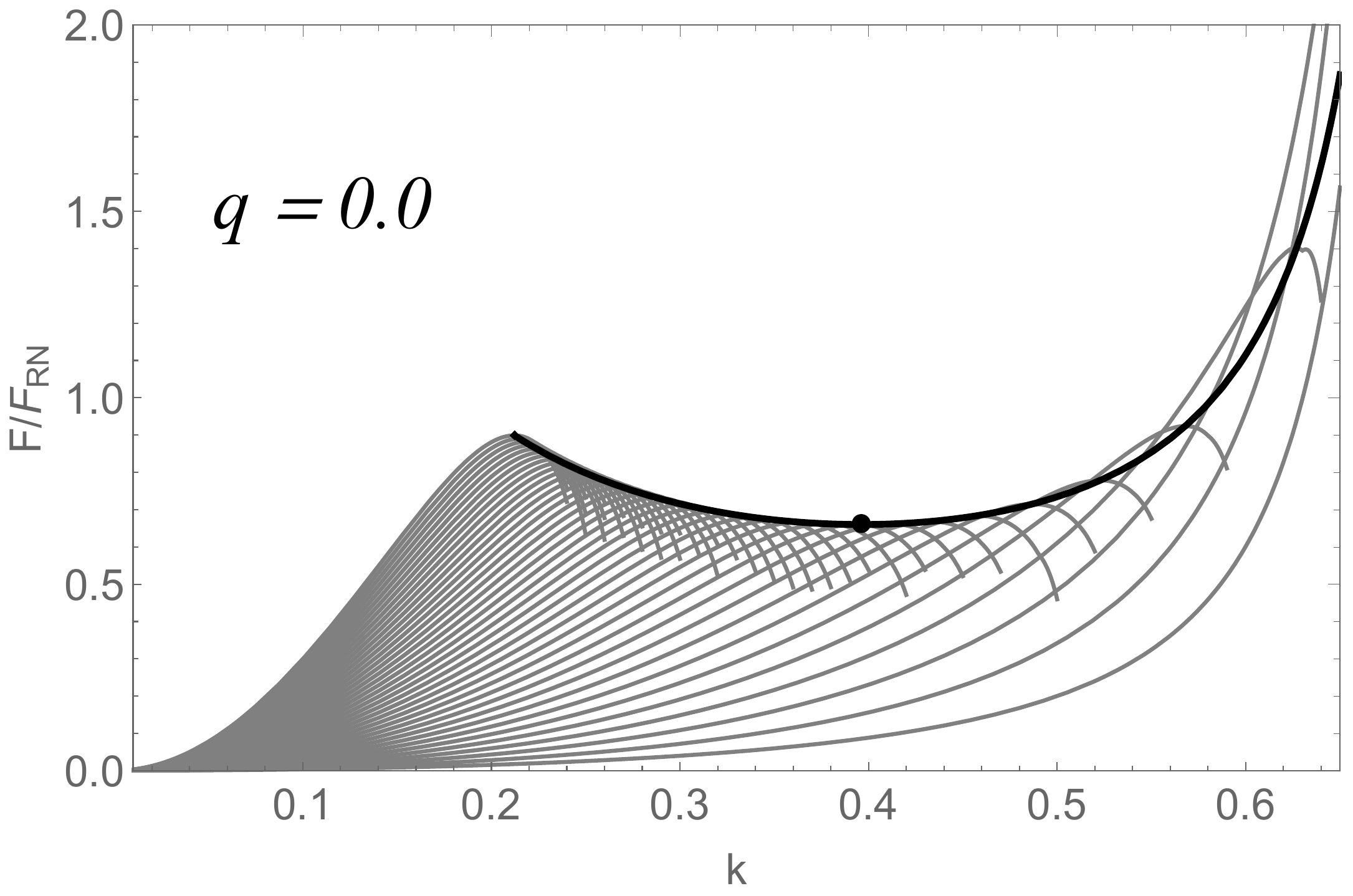}
    \includegraphics[scale=0.4]{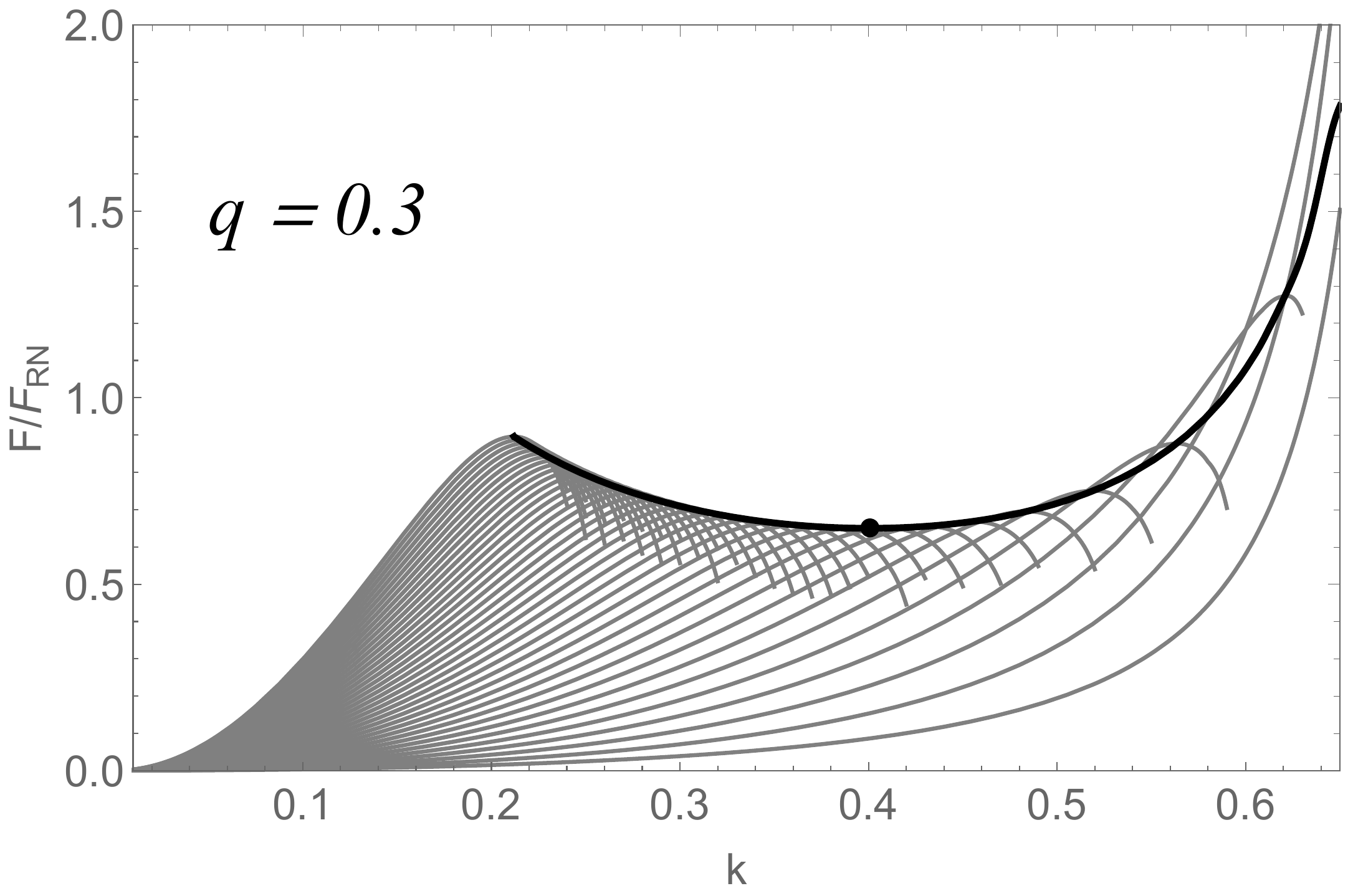}
    \includegraphics[scale=0.4]{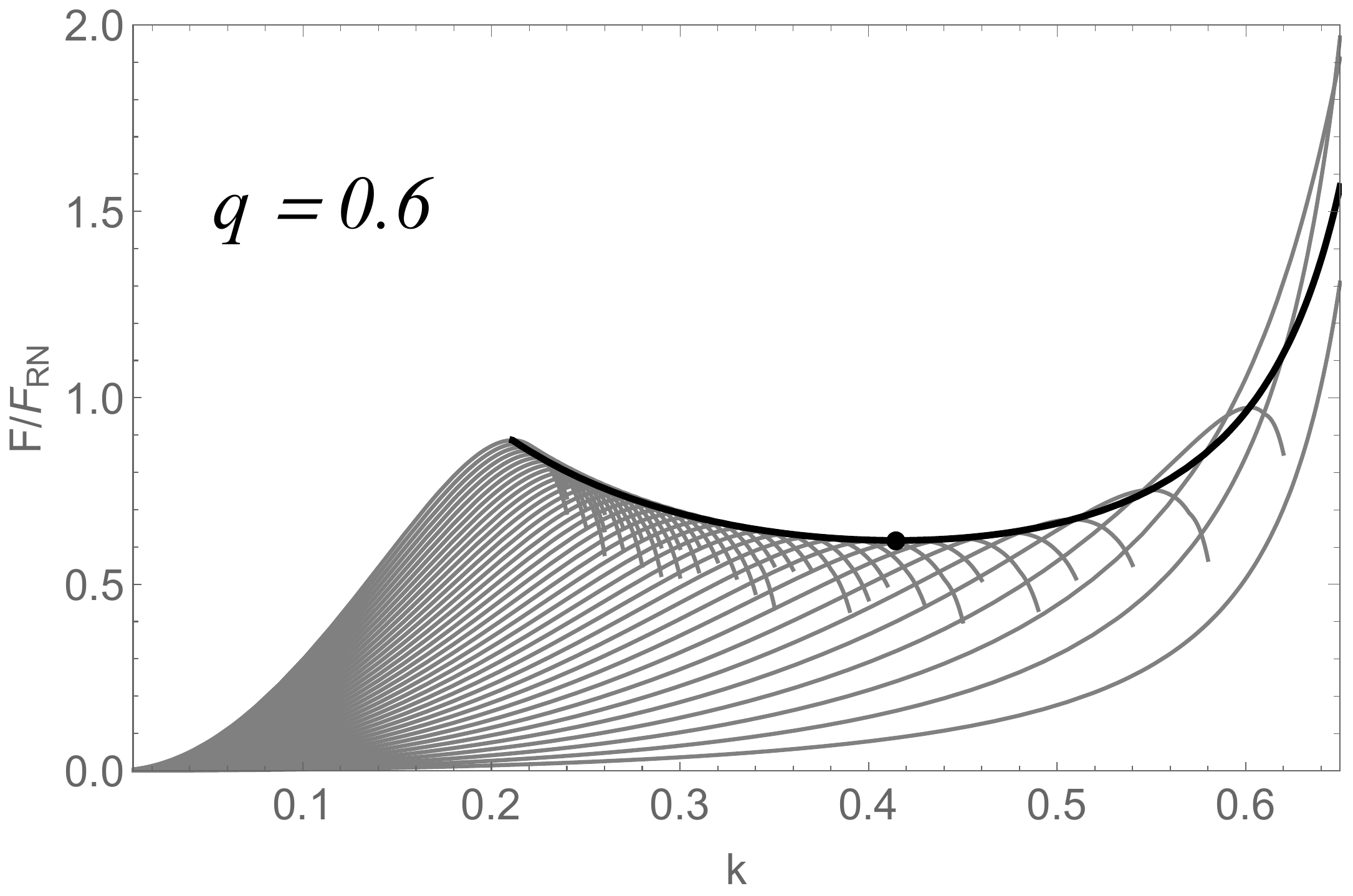}
    \includegraphics[scale=0.4]{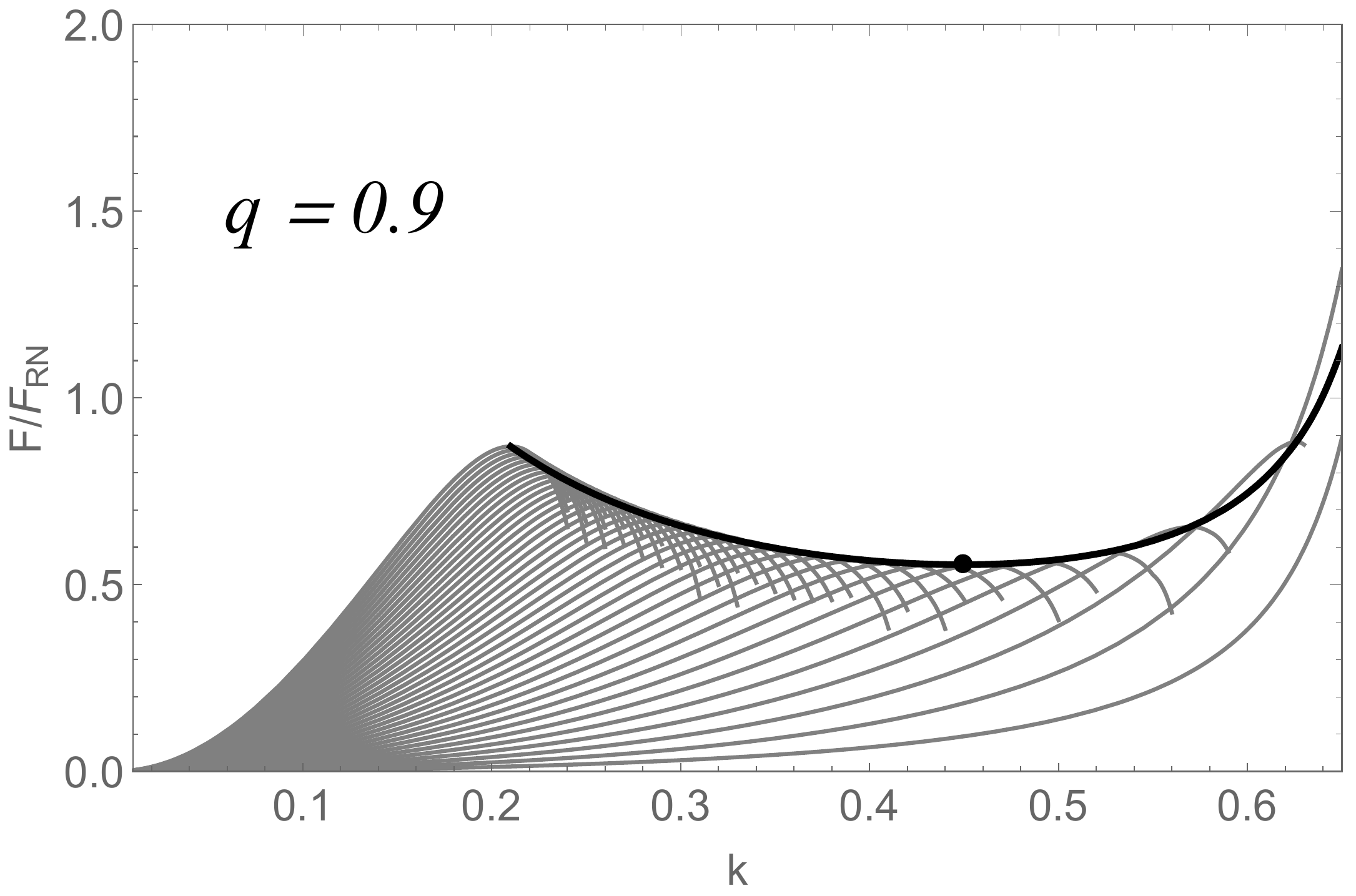}
    \caption{Normalized flux $F(k,x,q)/F_{RN}$ as a function of modulus $k$, for multiple values of the radial coordinate $x \in ( 1, 20 ] $, as well as discrete values of the charge-mass ratio $q$ as indicated. The dots on the curves represent the corresponding values of $F_\star$ and $k_c$ at which the accretion flow is stable.}
    \label{representacion k}
\end{figure*}

Before describing the general procedure for numerically solving the function $F(k,x,q)$, we define the physical range of interest as follows:  $k \in[0,1/\sqrt{2})$, $x \geq 1$ and $ q \in [0,1]$. For the latter, we focus mainly on fixed charge-mass ratio values when computing the energy flux. It should be noted that the extremal case $q=1$ was included for illustrative purposes only, as the true range of interest is $q \in [0,0.9]$ based on the current observational constraints provided by the EHT collaboration \cite{EventHorizonTelescope:2021dqv,EventHorizonTelescope:2022xqj,Vagnozzi:2022moj}.
Once $q$ is fixed, our procedure consists in varying simultaneously the remaining two parameters $ k $ and $ x $. This will allow us to determine the function $k(x)$ which
is at this point of the analysis unknown. Furthermore, it is important to note that not all values of $x$ lead to suitable values of $k$ in the range $[0,1/\sqrt{2})$. When this occurs, the numerical routine is stopped. Specifically, each value of $x$ (and $q$) is associated with a range of $k$ values and a maximum value of the energy flux $F(k,x,q)$. This issue of convergence is always under control and will be illustrated in more details later. Unlike Ref.~\cite{Brax:2019npi}, we explore a larger number of discrete values for $k$ and $x$ within the aforementioned ranges, resulting in a more detailed picture of the behavior of the energy flux.

In Fig. \ref{representacion k}, we represent the normalized flux $F/F_{RN}$ as a function of $k$, for various values of the radial coordinate $x \in (1,20]$. It is essential to keep in mind that $k$ scales inversely with $x$, meaning that smaller values of $x$ allow for a wider range of $k$ values to be covered. For example, among all sequences of curves, the one with the greatest spread corresponds to $x=1$, which diverges near $k\approx 0.7$ due to the singularity of coordinates\footnote{This divergence is associated with a coordinate singularity and can be observed from the square root argument in Eq. (\ref{Flujo final}). Additionally, it should be noted that there is a change of sign for certain values of $x$ that would result in $k\geq 0.7$. However, this situation never arises as these values lie beyond the range of interest. Consequently, the square root argument can be negative, thereby constraining the permissible values of $x$.}. Therefore, we will only consider curves that exhibit a maximum for the correct physical interpretation. This is the first inference drawn from  numerical analysis. Notice that this feature changes slightly as $q$ increases, as can be appreciated from the other panels.
 
As we approach both smaller and larger values of $x$, we observe an increase in $F/F_{RN}$. This behavior occurs as we move away from a critical point that represents the minimum of all possible maximum fluxes. To better illustrate this trend, we display an envelope made up of maximum fluxes, with a black point indicating the minimum flux. Another qualitative feature to note is that the normalized flux grows as $x^2$, as shown in Eq. (\ref{Flujo final}). In fact, for $x \gg 1$ the condition $ k \ll 1$ must be guaranteed
to obtain physical solutions.

The introduction of a BH charge $q$ has intriguing implications for both the energy flux and the stable accretion flux. As $q$ increases, all sequences of curves shift slightly toward the right, leading to a corresponding shift in the critical point (minimum of the envelope) toward the right. This effect is more pronounced when comparing the cases $q=0.3$ and $q=0.9$, where all curves are more tightly packed, resulting in a smaller region below the envelope. This suggests that the normalized flux is reduced
due to the presence of the charge $q$. In physical terms, an increase in $q$ directly impacts the size of the horizon (as shown in Eq. (\ref{new horizont})), making it smaller. Consequently, the accretion flux decreases since the closed surface at which the energy flux passes gets smaller. Therefore, our results are consistent with what is expected from accretion onto a charged BH.

\begin{figure*}
    \centering
    \includegraphics[scale=0.4]{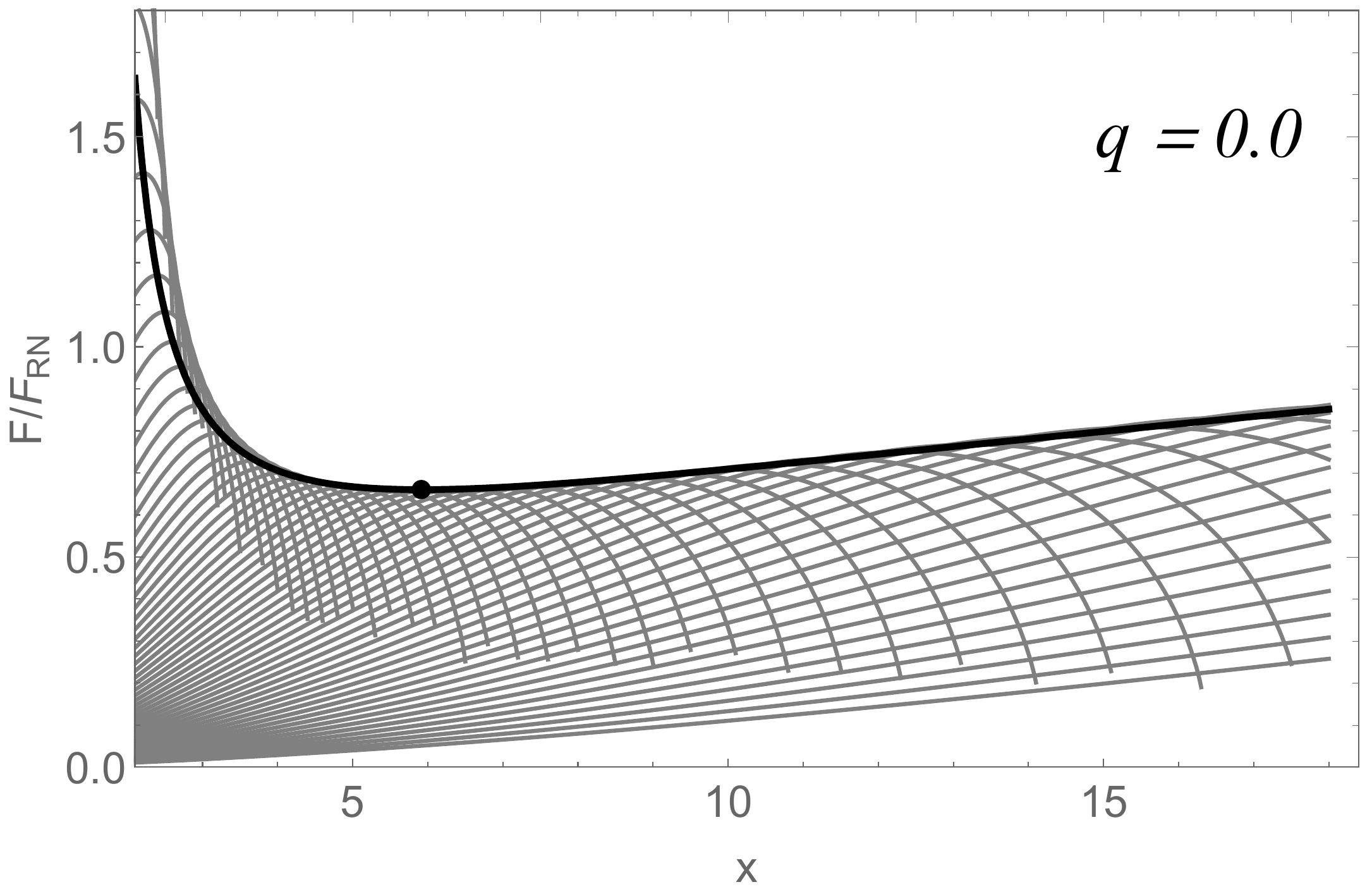}
    \includegraphics[scale=0.4]{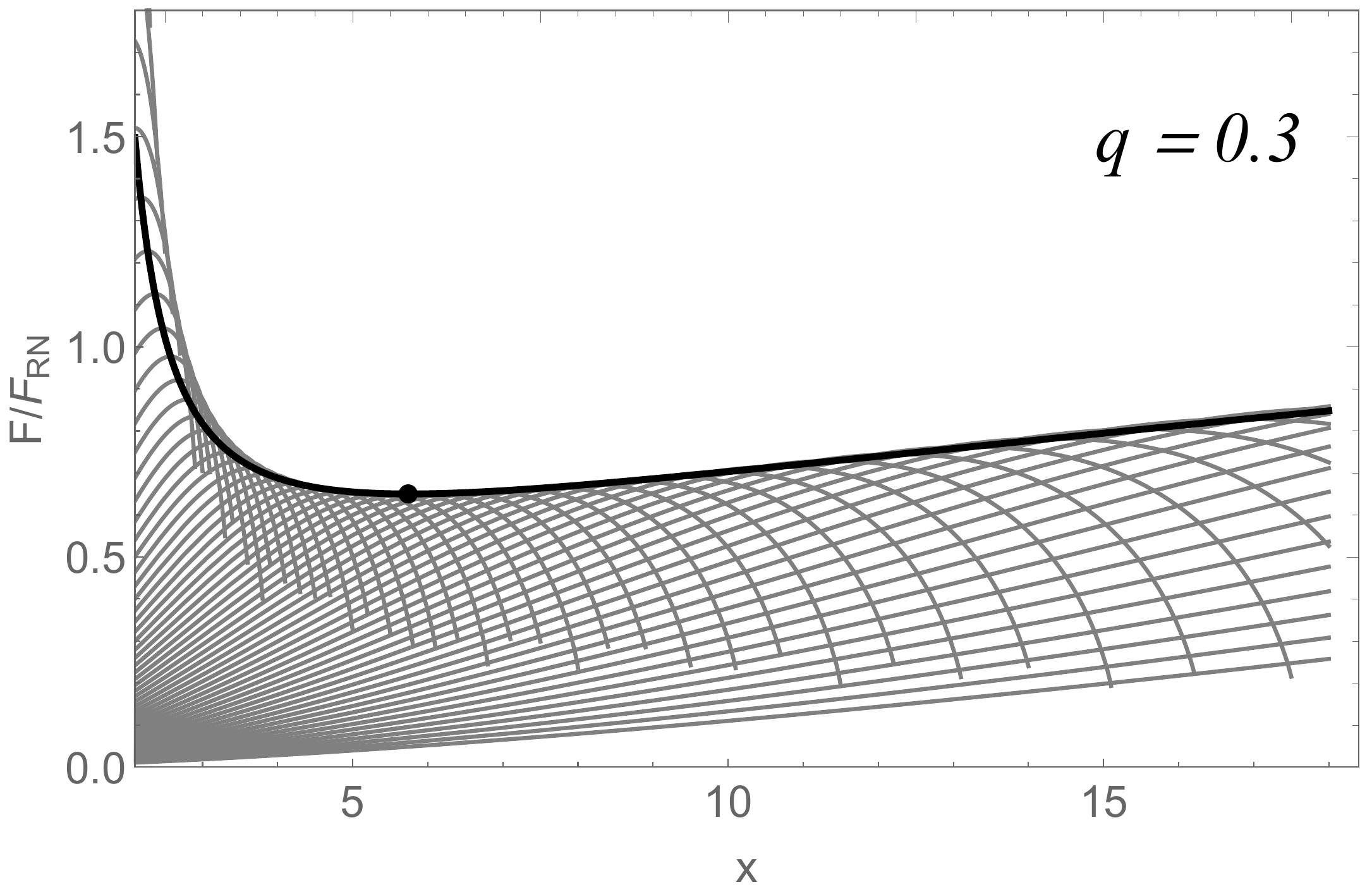}
    \includegraphics[scale=0.4]{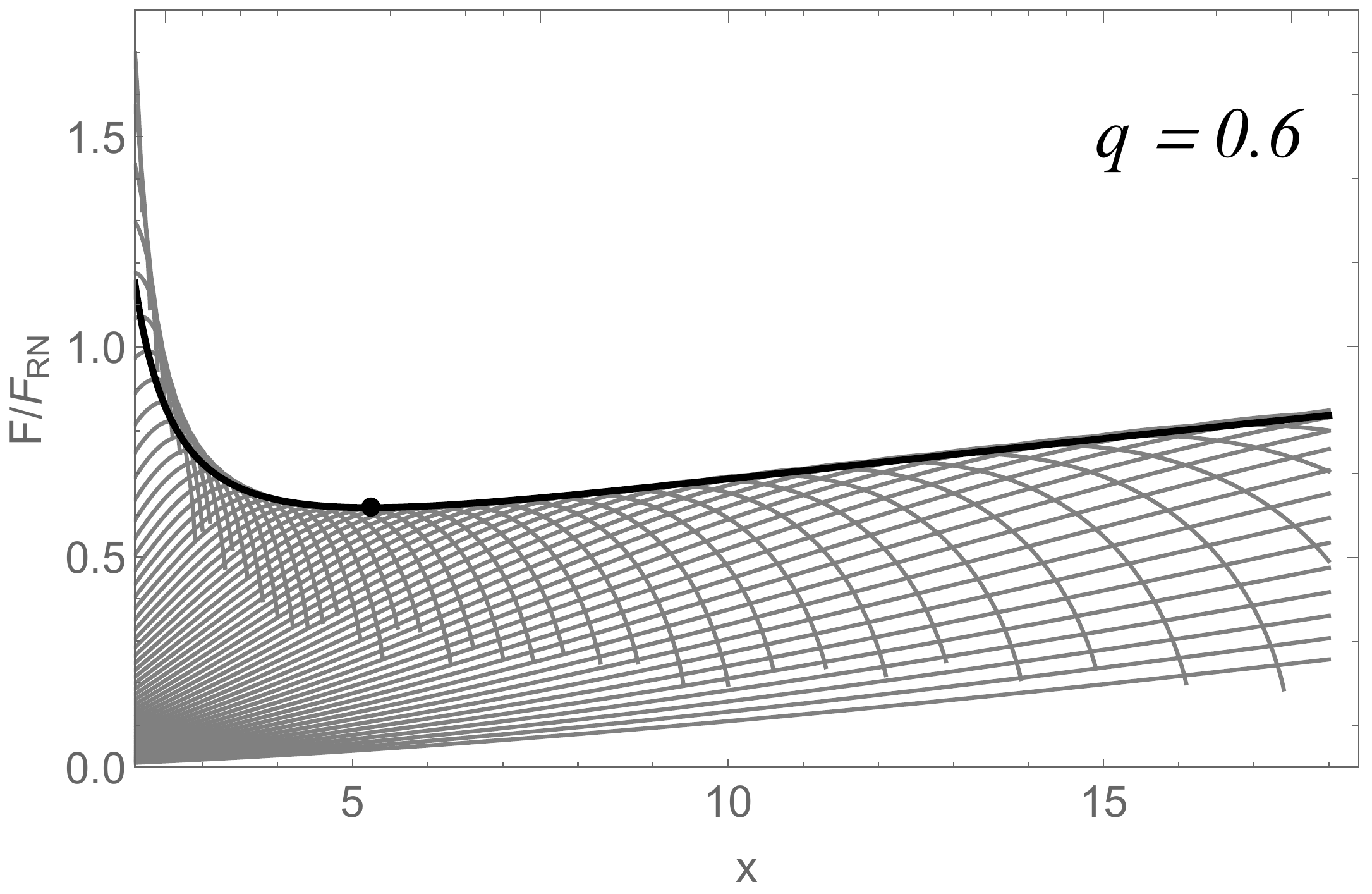}
    \includegraphics[scale=0.4]{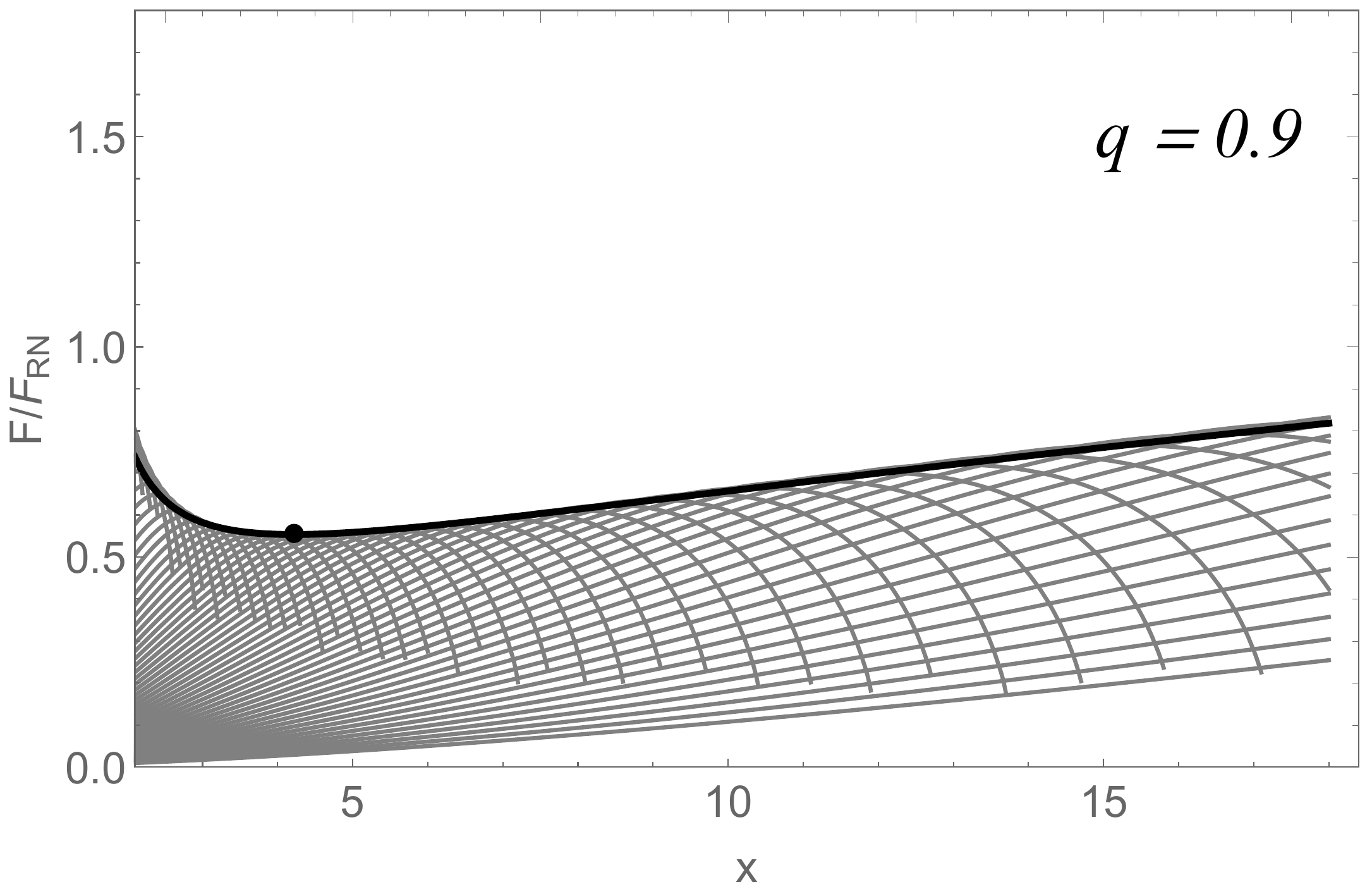}
     \caption{Normalized flux $F(k,x,q)/F_{RN}$ as a function of the radial coordinate $x$, for multiple values of $k$  $\in [ 0, 1/\sqrt{2}) $, as well as discrete values of the charge-mass ratio $q$ as indicated. The dots on the curves represent the corresponding values of $F_\star$ and $x_\star$ at which the accretion flow is stable.}
    \label{representacion x}
\end{figure*}

All of the aforementioned features can also be evidenced by plotting the normalized flux as a function of the radial coordinate $x$ for different values of $k$, as shown in Fig. \ref{representacion x}. The resulting curves satisfy the condition that for larger $x$ values, $k$ must be small and vice versa. The envelope is also constructed from the maximum fluxes, with a black point indicating the minimum flux. It is more noticeable in this representation that as $q$ increases, all energy fluxes decrease. As a result, the critical flux becomes smaller due to the smaller size of the horizon. In addition, the values of $x$ associated with the critical flux become smaller as well.

\begin{figure*}
    \centering
\includegraphics[scale=0.4]{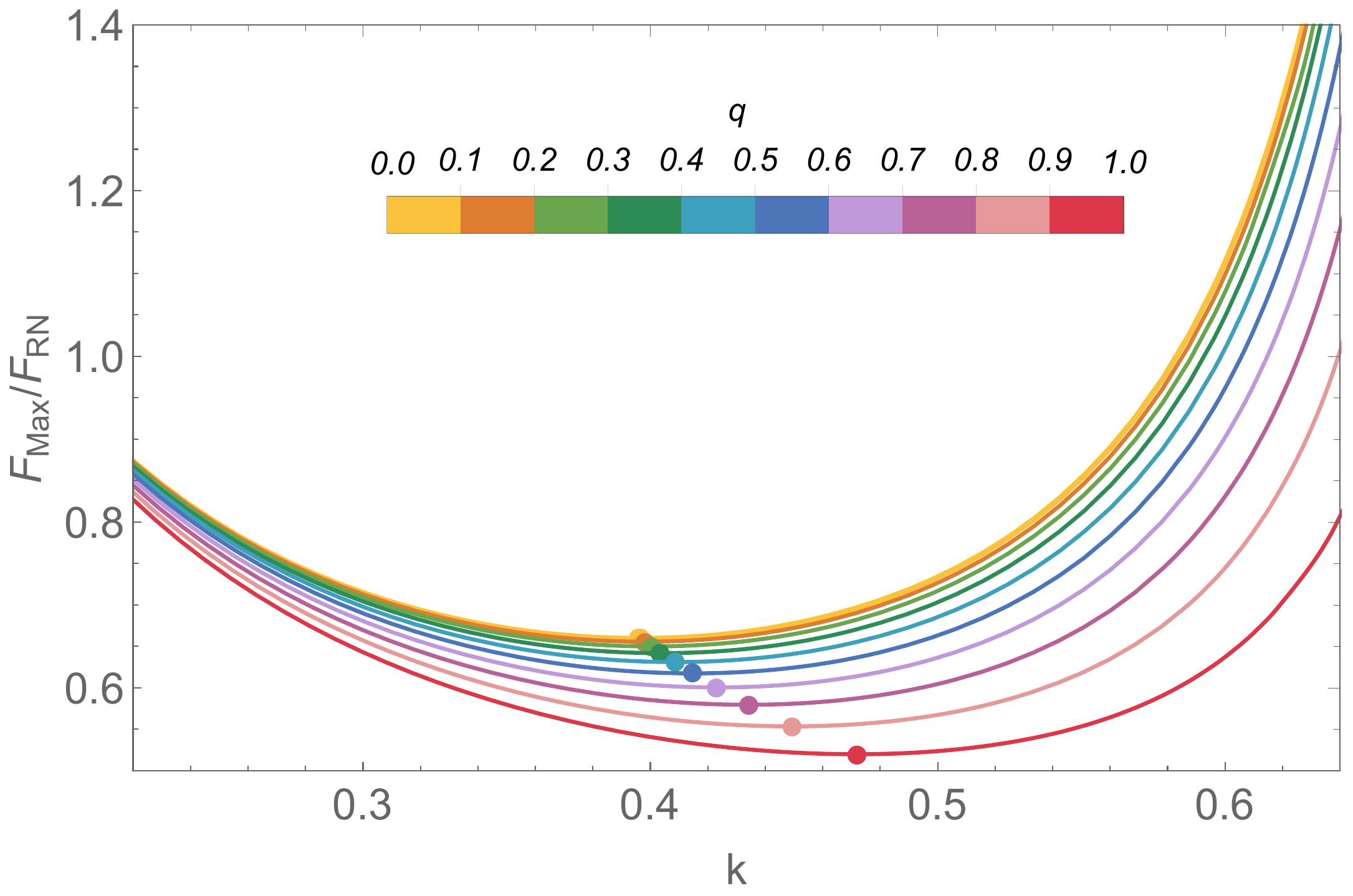}
\includegraphics[scale=0.4]{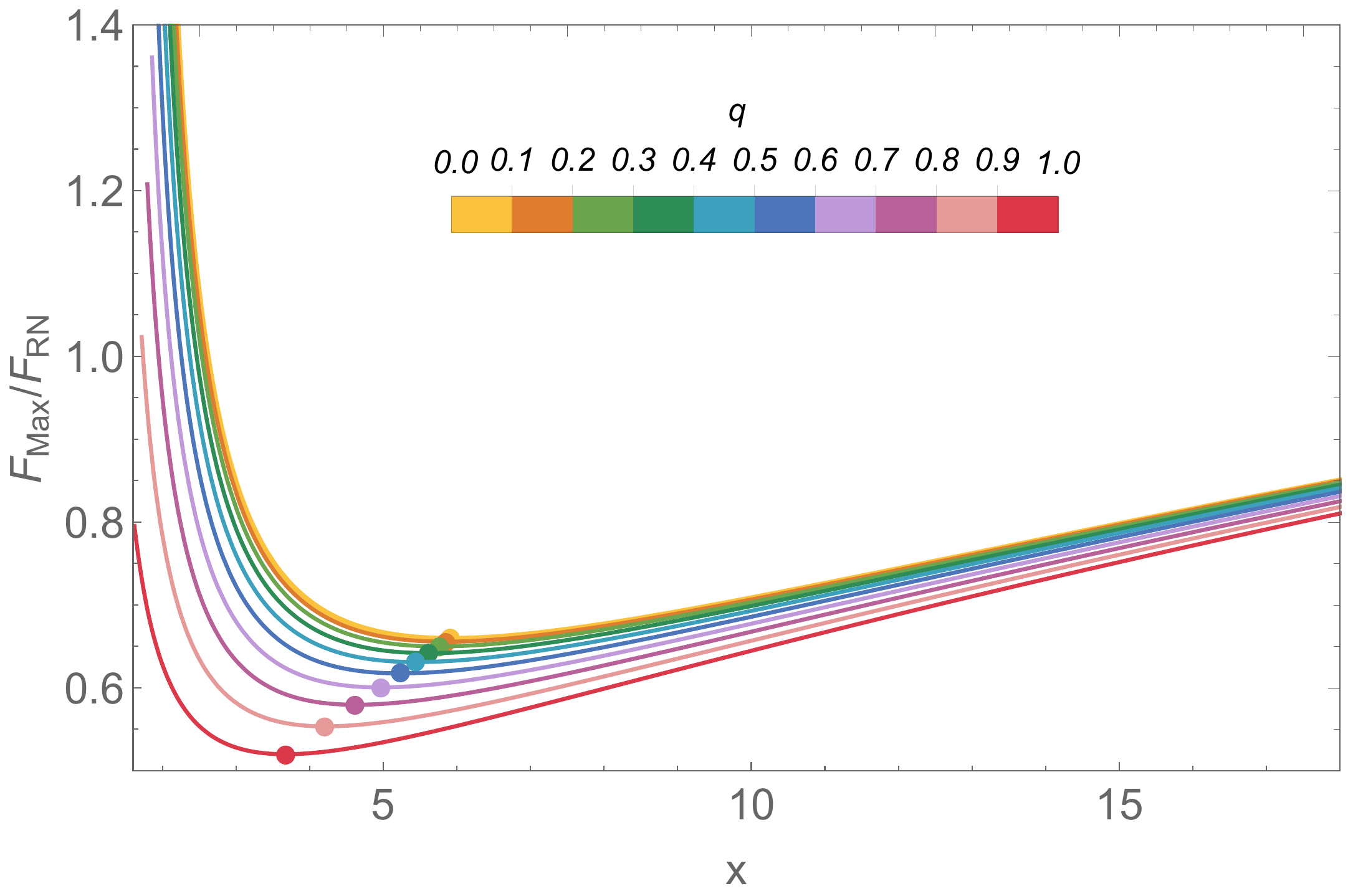}
\caption{Left and right panels display the same envelopes of Figs. \ref{representacion k} and \ref{representacion x} as a function of $k$ and $x$, respectively, for various charge-mass ratios $q$ $\in [0, 1]$, the upper curve in yellow represents $q=0$, and as we move down the curves, the value of $q$ increases until reaching $q=1$ shown in dark red. These curves are constructed from the maximum values of the energy flux. The dots on the curves represent the critical values of $F_\star $, $k_c$, and $x_\star $.}
    \label{envolventes}
\end{figure*}

It is instructive to plot all the envelopes in a single figure for different charge values, either as a function of $k$ (left panel of Fig. \ref{envolventes}) or as a function of $x$ (right panel of Fig. \ref{envolventes})\footnote{The envelopes of Fig. \ref{envolventes} were constructed from the maximums energy fluxes obtained simultaneously from Fig. \ref{representacion k} and Fig. \ref{representacion x}.}. The corresponding critical points are denoted by points of different colors, representing distinct charge values, on their respective curves. We display, however, charge values ranging from $0$ to $1$ for completeness. From Fig. \ref{envolventes}, it is clear how the energy flux decreases as $q$ increases. Furthermore, $k$ becomes larger while $x$ becomes smaller. In other words, the critical point moves from above to below as $q$ increases.

As the main conclusion of this numerical analysis, we observe then the existence of different inflection points for different values of $q$, which correspond to a critical flow $F_c$. This critical flow shows a minimum in $k_c$ and $x_\star$ for a particular $q$. Thus, as $q$ increases, the value of $F_c$ decreases, causing $k_c$ shifts toward higher, while $x_\star$ moves toward lower values. The decrease of $F_c$ and $x_\star$ as $q$ increases is the result of the explicit dependence on the metric function $f(x,q)$ and the scaling $F\sim x^2$.

It is convenient to define, similar to the Schwarzschild case \cite{Brax:2019npi}, the following expression for the critical flux
 \begin{equation}
     F_c = F_{\rm {Max}}(x_\star) = F_\star F_{RN}.\label{F critico}
 \end{equation}
In this way, we can group all the relevant information $F_\star $, $k_c$, $x_\star $ and $q$ from Fig. \ref{envolventes} into Table \ref{resumen}. From the table we can see that $x_\star$ is located in the region known as marginally bound orbit $r_{\rm {mb}}$, which is between the photon sphere $r_{\rm {ph}}$ and the Innermost stable circular orbit $r_{\rm {isco}}$, that is, $ r_{\rm {ph}} < r_{\rm {mb}} < r_{\rm {isco}}$. In this region, it is inevitable that the SF falls into the RN-BH. This is the reason why the maximum flow occurs in this region, which is consistent with the findings reported in \cite{Feng:2021qkj,Richards:2021zbr}. It is important to mention that the particular case $F_\star = 0.66$ corresponds to the Schwarzschild BH  studied in \cite{Brax:2019npi}.

On the other hand, we can see that if $F/F_{RN} < F_c$, there exist two solutions of $k(x)$: $k_1(x)$ and $k_2(x)$. That is, there are two $x$ values that provide the same $F/F_{RN}$. On the contrary, if $F/F_{RN} > F_c$, there is no solution of $k(x)$, as can be verified by examining Eq. (\ref{Flujo final}). When $F/F_{RN} = F_c$, the solution demands $k_1(x) = k_2(x)$. This equality takes place at $x_\star$ and $k_c$, which means that for a stable critical energy flux to exist, there must be a smooth and continuous transition from low-velocity (large radii) to high-velocity (radii close to the horizon) \cite{Brax:2019npi,Brax:2020tuk,Boudon:2022dxi,Feng:2022bst,Boudon:2023qbu}. This is the same criterion used to have transonic solutions in the hydrodynamic case of polytropic infall fluid onto a BH \cite{Bondi:1952ni}. 

\begin{figure}
    \centering
\includegraphics[scale=0.41]{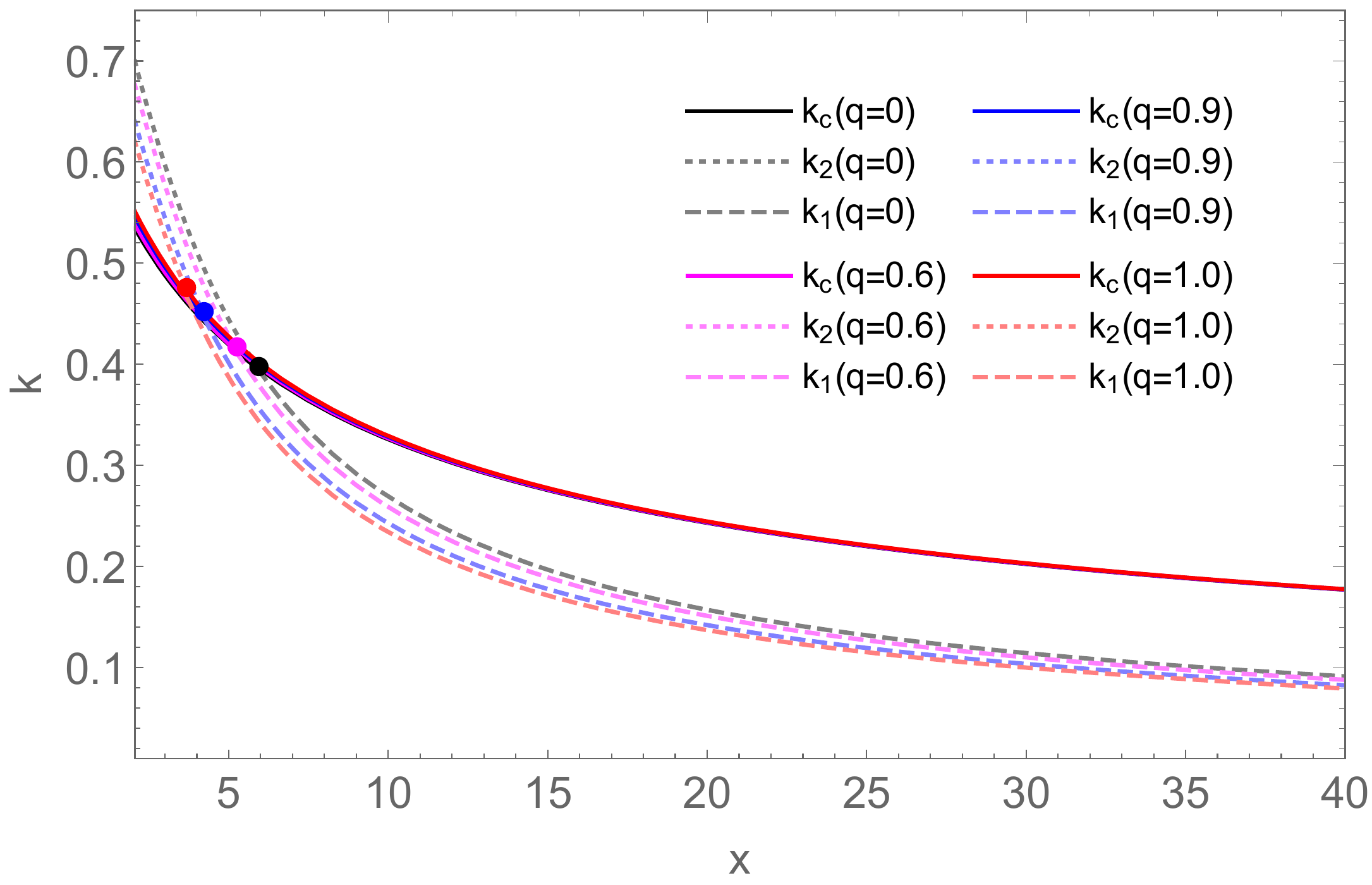}
\caption{Modulus $k_{1}$ (dotted lines), $k_2$ (dashed lines) and $k_{c}$ (solid lines) as a function of the radial coordinate for different flows $F_c$ and associated $q$. Dots on the curves indicate the change from the branch $k_{1}(x)$ to $k_{2}(x)$ with associated ($x_{\star}$,$k_{c}$) coordinates.}
    \label{Relacion k vs x}
\end{figure}

Fig. \ref{Relacion k vs x} depicts different curves of $ k_1(x)$, $k_2(x)$, and $ k_c(x) $ for various  values of $q$ and their corresponding critical energy fluxes $F_c$. The solution $ k_2(x)$ describes the low-velocity regime, while the solution $ k_1(x)$ describes the high-velocity regime. In addition, all points denote the $x_\star$ at which the transition between the two solutions occurs, as $k_c(x)$ is continuous.  The curve $k_c$ is constructed from the branches mentioned above as follows: If $x < x_\star $, then $k_1(x)= k_c(x) $, and if $x > x_\star $, then $k_2(x)= k_c(x)$. The unexpected finding is that the transition points for different $q$ values lie on the same curve $k_c$, indicating that they are independent of the charge-mass ratio. However, as $q$ increases, the transition point $x_\star$ shifts along the same $k_c$ curve.

\begin{figure}
    \centering
\includegraphics[scale=0.41]{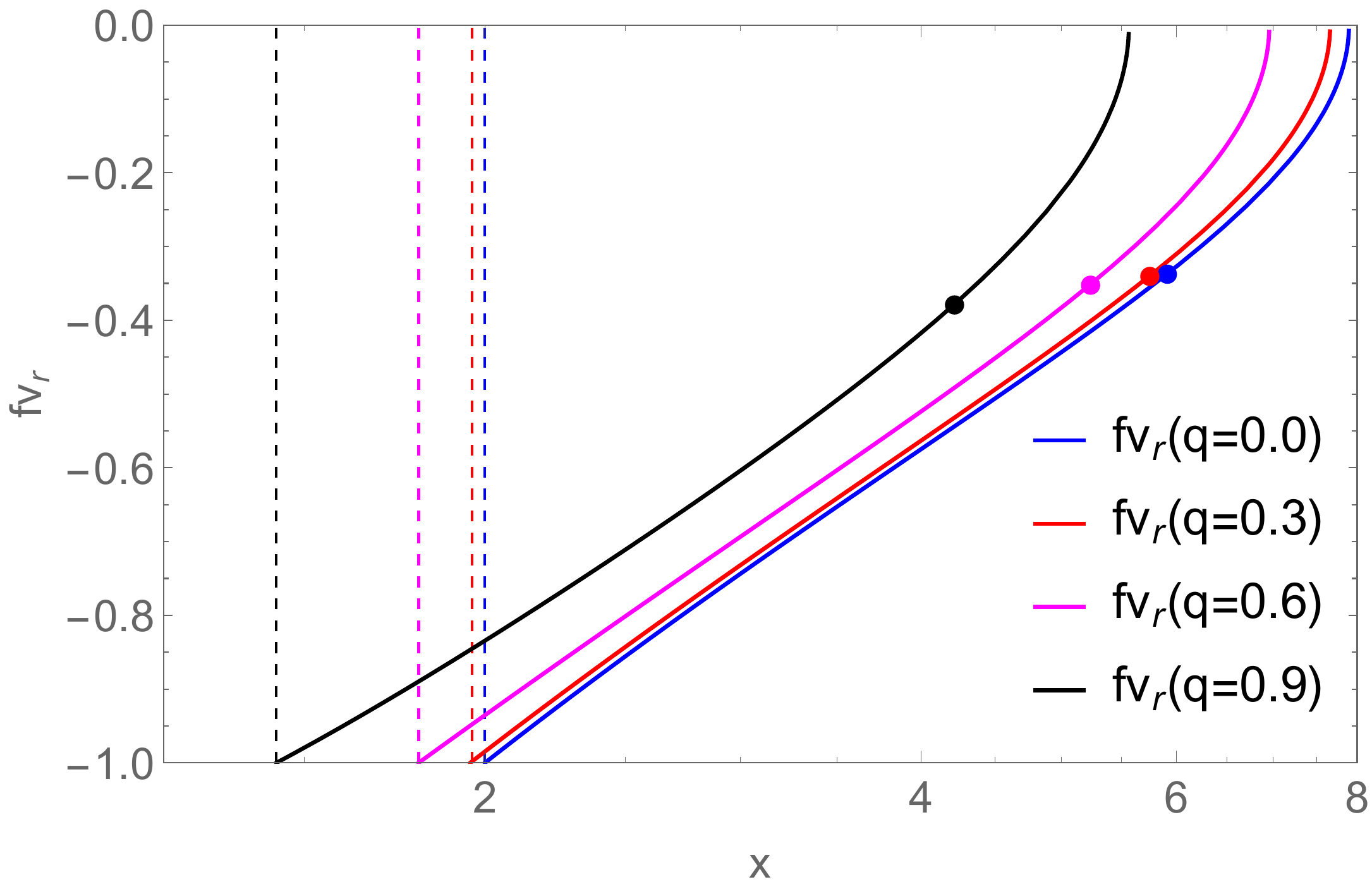}
\caption{Rescaled radial velocity $ f v_r(k_c,x,q)$ as a function of the radial coordinate $x$ from Eq. (\ref{velocidad}), for different modulus $k=k_c$ and discrete values of the charge-mass ratio $q$. Dots indicate the corresponding positions $x_\star$ and vertical dotted lines represent the corresponding horizons.}
    \label{velocity}
\end{figure}

For a clearer representation of the behavior of the infall velocity, we plot the rescaled radial velocity $fv_r (k_c,x,q)$ from Eq. (\ref{velocidad}) in Fig. \ref{velocity}. The plot shows the behavior of $fv_r$ as a function of $x$ for different values of $q$ and $k=k_c$. The vertical dotted lines indicate the horizons for different values of $q$. The points on these curves represent the distance $x_\star$ at which the flow becomes transonic, allowing the change from the low-velocity branch to the high-velocity branch. Moreover, it is worth noting that near the horizon, the flow velocity becomes ultrarelativistic \cite{Papadopoulos:1998up}. The effect of $q$ is notable: as it increases, the size of the horizon decreases, bringing the critical point closer to it. However, the value of the critical velocity does not change significantly\footnote{The same feature can be observed in the accretion process of a polytropic fluid around a RN BH, as described, for instance, in \cite{Babichev:2008jb,Gomez:2023wei}.}. As the flow passes through the critical point, the steady infall of the SF onto the BH takes place, which determines the resulting SF profile near the BH, as in the hydrodynamical case \cite{Shapiro:1983du}. Finally, it is worth noting that far beyond the gravitational influence of the BH, the accretion process ceases due to the hydrostatic equilibrium condition.

\begin{table}
\begin{center}
\begin{tabular}{|c|c|c|c|}
\hline
 $q$ & $F_\star $ & $k_c$ & $x_\star $ \\
 \hline
 0. & 0.66013  & 0.39606 & 5.91908 \\
 \hline
 0.1 & 0.65902 & 0.39659 & 5.90113 \\
 \hline
 0.2 & 0.65564 & 0.39829 & 5.84691 \\
 \hline
 0.3 & 0.64994 & 0.40049 & 5.7510 \\
 \hline
 0.4 & 0.64183 & 0.40325 & 5.62122 \\
 \hline
 0.5 & 0.63107 & 0.40885 & 5.45029 \\
 \hline
 0.6 & 0.61744 & 0.41483 & 5.23017 \\
 \hline
 0.7 & 0.60048 & 0.4228 & 4.95924 \\
 \hline
 0.8 & 0.57953 & 0.43392 & 4.62737 \\
 \hline
 0.9 & 0.55346 & 0.44903 & 4.21467 \\
 \hline
 1. & 0.51983 & 0.47158 & 3.6691 \\
 \hline
\end{tabular}
\end{center}
\caption{\label{resumen} Summary of critical values $F_\star $, $k_c$ and $x_\star $ for different values of $q$ extracted from Fig. \ref{envolventes}. The cases $q=0$ and $q=1$ correspond, respectively, to the Schwarszchild and extremal BHs.}
\end{table}

\subsection{Density profile}

To obtain a complete picture of the accretion process, it is necessary to provide the energy density. In the hydrodynamical case, it can be derived from the relativistic version of the Bernoulli equation (see, e.g., \cite{Shapiro:1983du}). Using Eqs. (\ref{Solucion K-G no lineal}) and (\ref{rho_scalar}), and considering the relations given by Eqs. (\ref{relacion temporal phi}) and (\ref{relacion radial phi}), we can compute the energy density for the RN metric in the large scalar mass limit, yielding the following result
\begin{widetext}
    \begin{equation}
       \rho_\phi = \frac{m^4}{\lambda_4} \frac{k^2}{(1-2 k^2)}\left[\frac{1}{f}( 1 - k^2 + (2 k^2-1){\rm cn}^2 - k^2 {\rm cn}^4)\left(\frac{2 {\bf K} (1+\alpha)}{\pi}\right)^2\left(2-\frac{\pi^2 f}{(1-2 k^2) 4 {\bf K}^2 (1+\alpha)^2}\right)+ {\rm cn}^2 + \frac{k^2{\rm cn}^4}{(1-2 k^2)}\right].\label{Perfil densidad Schwarzschild} 
    \end{equation}
\end{widetext}
We can express the previous equation in terms of the flow $F_c$, by using Eq. (\ref{definicion F_RN}) and averaging the fast oscillations over time. This enables us to obtain the following

\begin{widetext}
    \begin{align}
       \langle  \rho_\phi \rangle &= - \frac{F_c}{F_\star (2M)^2}\frac{k^2}{(1-2 k^2)}[\frac{1}{f}(1 - k^2 + (2 k^2-1) C_2 - k^2 C_4 )\left(\frac{2\bf K}{\pi}\right)^2\left(2-\frac{\pi^2 f}{(1-2 k^2) 4 {\bf K}^2 (1+\alpha)^2}\right)\nonumber\\ &+ \frac{1}{(1+\alpha)^2}\left(C_2 + \frac{k^2 C_4}{(1-2 k^2)}\right)].\label{promedio Perfil densidad Schwarzschild}
    \end{align}
\end{widetext}

We can observe that the average energy density of the scalar field, $\langle \rho_\phi \rangle$, diverges as $x$ approaches the horizon due to the $1/f$ term. However, this singularity is an artifact of using nonregular coordinates at the horizon. Setting the self-interaction to zero can apparently pose problems due to the divergence in Eq.~(\ref{Perfil densidad Schwarzschild}). However, as $\lambda_{4}$ approaches zero, $k$ must also strictly do so. Therefore, we use this condition to deduce the corresponding density profile for the scalar free case, ensuring thus consistency. This is shown in the appendix.
Additionally, we can see that for $x \gg 1$, $ \langle \rho_{\phi} \rangle \propto r^{-1} $, in contrast to the noninteracting case where $ \langle \rho_{\phi} \rangle \propto r^{-3/2} $ \cite{Hui:2019aqm,Brax:2019npi}.

\begin{figure}
    \centering
    \includegraphics[scale=0.42]{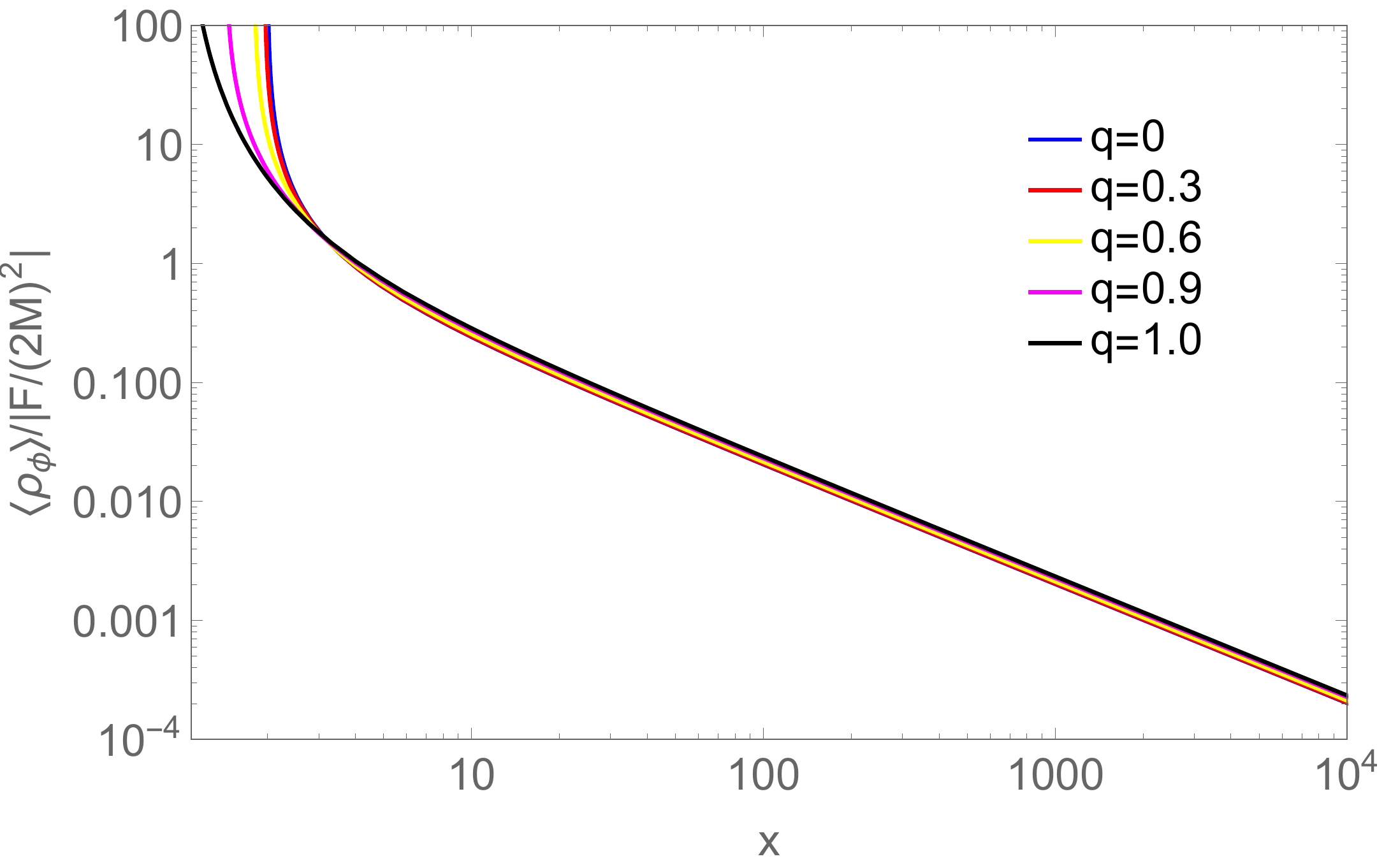}
    \caption{Normalized energy density of the scalar field for different charge-mass ratios $q$, as given by Eq. (\ref{promedio Perfil densidad Schwarzschild}). Here $x$ ranges from a position close the horizon to $10^4$ times the horizon. The abrupt increase of the density around the corresponding horizon is due to the use of nonregular coordinates at the horizon. For illustrative purposes, we have included the case $q=1$.}
    \label{Densidad de energia campo escalar}
\end{figure}

Fig. \ref{Densidad de energia campo escalar} shows the normalized energy density of the scalar field $\langle\rho_\phi \rangle / |F/(2M)^2|$ as a function of $x$ for different values of $q$. The figure illustrates the aforementioned divergences when $x$ approaches the corresponding horizon, and makes more evident the effect of changing $q$ on the density profile. When $x\gg 1$, we have $k\ll 1$, and therefore $\langle \rho_\phi \rangle / |F/(2M)^2|\propto x^{-1}$. It is also appreciable that a variation in $q$ produces a slight difference between the energy densities at large distances, as expected. An increment in $q$ leads to a reduced horizon, consequently diminishing the region where self-interacting SF infalls. This, in turn, directly impacts the density profile. By using regular coordinates as in \cite{Brax:2019npi}, it can be shown that the energy density becomes constant at small radii where self-interaction becomes subdominant.

\subsection{Accretion}
 
\begin{figure}
    \centering
\includegraphics[scale=0.45]{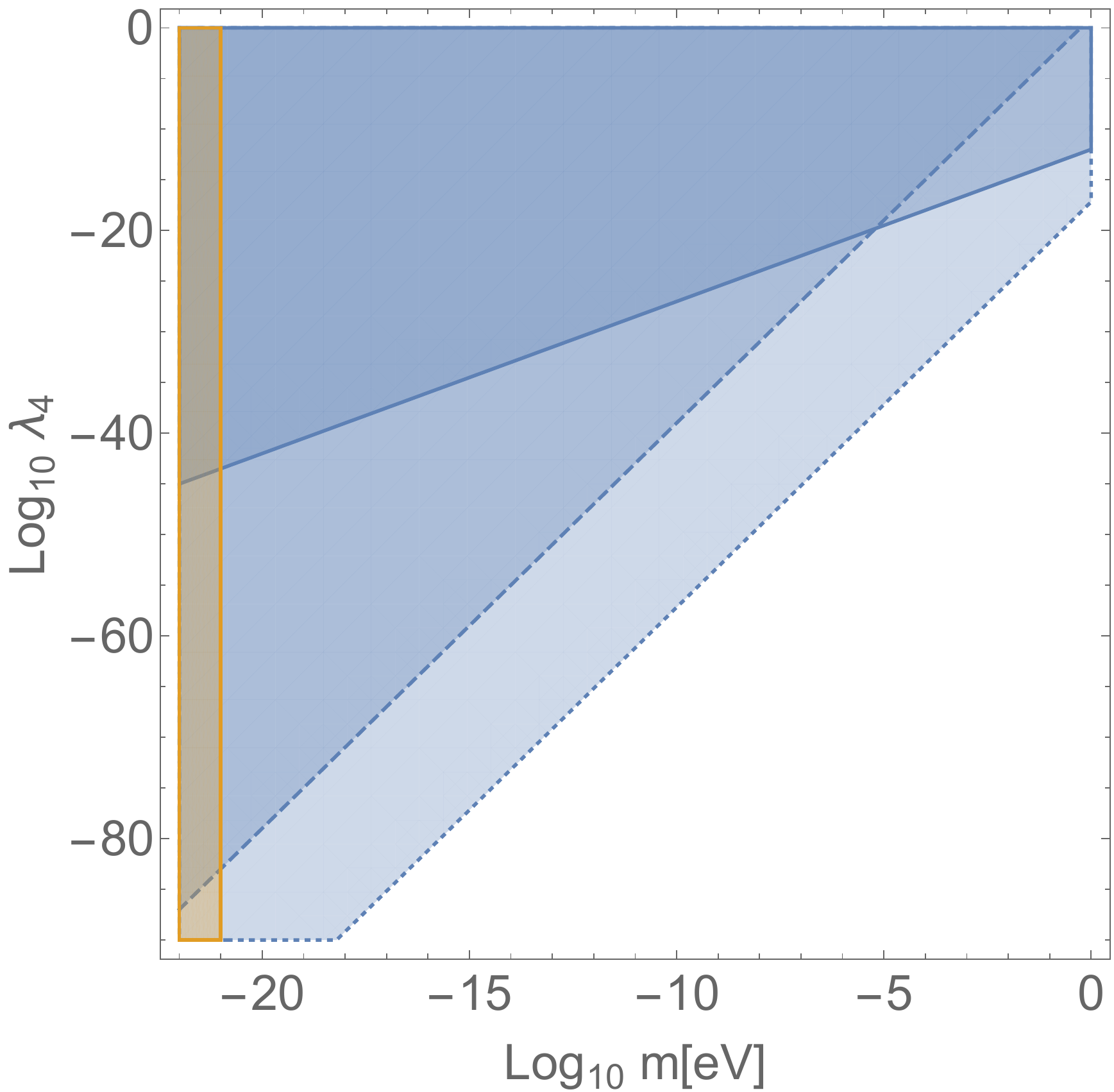}
\caption{Exclusion regions for the parameter space ($m$,$\lambda_{4}$) using different constraints.  When quantum pressure effects are neglected (shaded vertical area), observation of galaxy cluster merging (solid diagonal line), changes in the propagation speed of GWs (dotted diagonal line), and finally, our constraint due to accretion in M87$^{\star}$ Eq. (\ref{constrain ecuacion}) (dashed diagonal line).}
    \label{constrain lambda y masa}
\end{figure}

In this section, we compute the mass accretion rate around a RN-BH. The mass accretion can be obtained from the energy-momentum tensor equation $\nabla_\mu T^\mu_\nu = 0$. For a steady state, the mass accretion rate is defined as the energy flow through a closed surface of a sphere, given by
\begin{equation}
    \dot M(r) =   \oint T^r_t\sqrt{-g}d\theta d\varphi.\label{definicion de M punto}
\end{equation}
Considering the critical flux $F_\star(q)$ defined in the previous section [Eq. (\ref{F critico})] and compiled in Table \ref{resumen} for different $q$ values, the mass accretion rate can be expressed as
\begin{equation}
    \dot M_{\rm SIDM} = 4\pi F_\star(q) \frac{(2M)^2 m^4}{\lambda_4},\label{acrecion DM}
\end{equation}
which is consistent with previous findings \cite{Brax:2019npi,Feng:2021qkj}.  Hence, from theoretical considerations, it is expected that DM may contribute to the mass accretion rate of BHs, $\dot{M}_{\rm BH}$, in addition to baryonic matter, such that
\begin{equation}
   \dot{M}_{\rm BH} = \dot{M}_{\rm baryons} + \dot{M}_{\rm DM}. \label{BHaccretion}
\end{equation}
A typical model of steady-state accretion of baryons establishes a well-known linear relation between luminosity $L$ and accretion mass $\dot M_{\rm Baryons}$. From this relation, the observed luminosity is related to the hot gas where the infalling kinetic energy is converted into heat and radiation. Therefore, it is reasonable to consider that the mass accretion rate of BHs is mostly accounted for by the mass accretion rate of baryons, i.e. $\dot{M}_{\rm BH} \sim \dot{M}_{\rm baryons}$. Hence $\dot M_{\rm baryons} \gg \dot M_{\rm DM}$. However, we consider using the relation given by Eq.~(\ref{BHaccretion}) to connect the observed mass accretion rate with the one given by dark matter by imposing the condition that $\dot{M}_{\rm DM}$ cannot overcome $\dot{M}_{\rm baryons}$. In other words $\dot{M}_{\rm DM}\ll\dot{M}_{\rm baryons}$, but not negligible. From this physical condition, we can place a conservative bound on the parameter space by relating Eq. (\ref{acrecion DM}) with the recent measurement of BH accretion for M87$^{\star}$ obtained by polarization data of the EHT collaboration \cite{EventHorizonTelescope:2021srq}. The reported value for the mass accretion rate is $\dot M_{87} \sim$ (3-20) x $10^{-4} M_{\odot}$ yr$^{-1}$. Interestingly, this value allows us to obtain a lower limit for the self-interaction parameter for a given particle mass, which reads
\begin{equation}
    \lambda_4 > (1.49-10.2) \left( \frac{m}{1 \rm 
 eV}\right)^4 .\label{constrain ecuacion}
\end{equation}
In order to check the consistency of this result with other constraints, we present in Fig. \ref{constrain lambda y masa} the following constraints: a negligible quantum pressure for $ m \gg 10^{-21} \rm eV $ \cite{Brax:2019fzb}, represented by the vertically shaded area; observations of cluster mergers indicating that $ \lambda_4 \lesssim 10^{-12}(m/ 1\rm eV)^{3/2} $ \cite{Randall:2008ppe}, represented by the solid diagonal line; and the limit on the change in the speed propagation of gravitational waves (GWs) $\delta c_g \leq 10^{-20} $ \cite{Dev:2016hxv}, represented by the dotted diagonal lines. We also include the most conservative estimate of the accretion rate of M87$^\star$, $ \dot M_{87} \sim $ 3 x 10$^{-4} M_{\odot}$~yr$^{-1}$, which results in the smaller value of Eq. (\ref{constrain ecuacion}) and is represented by the dashed diagonal lines. To guarantee the validity of this result, we have to strictly ensure two important conditions: the mass-large limit and that for a particle mass, such as $m\sim 1$~eV, the quartic self-interaction $\lambda_{4}\ll1$. Even though we cannot improve the current bounds, the latter condition can be as competitive as, for instance, the one inferred by the change in the speed of propagation of GWs. This is very promising since future observations of the mass accretion rate of BHs may provide better constraints on the parameter space of the SF.

To gain a better understanding of the order of magnitude of the mass accretion rate given by Eq. (\ref{acrecion DM}), let us consider extreme values of $F_\star$, namely $F_\star(q=0)$ and $F_\star(q=1)$, along with typical values for a SF galactic halo properties, such as SMBH $\sim10^6M_\odot$, $m\sim10^{-5}$ eV, and $\lambda\sim10^{-19}$. For $q=0$ and $q=1$, these values yield self-interacting SF DM accretion rates of $\dot M_{\rm SIDM}\simeq8.15\times10^{-10}M_\odot~{\rm yr}^{-1}$ and $\dot M_{\rm SIDM}\simeq6.32\times10^{-10}M_\odot~{\rm yr}^{-1}$, respectively. Although there is a 22.5\% difference between the two accretion rates, both values remain on the same order of magnitude, $\dot M_{\rm SIDM}\sim10^{-10}M_\odot~{\rm yr}^{-1}$.

Comparing this theoretical result with the one reported in \cite{Feng:2021qkj}, where $\dot M_{\rm min}\sim1.41\times10^{-9}M_\odot~{\rm yr}^{-1}$, we can conclude that both results are of the same order of magnitude. This is also consistent with observed Eddington accretion rates of baryons, which are on the order of $\sim0.02M_\odot~{\rm yr}^{-1}$ \cite{Salpeter:1964kb}, Bondi accretion rates \cite{Bondi:1944jm,Bondi:1952ni,Edgar:2004mk}, and the accretion rate of the BH M87$^\star$, which is estimated to be $\dot M_{87}\sim(3-20)\times10^{-4}M_\odot~{\rm yr}^{-1}$ \cite{EventHorizonTelescope:2021srq}. In all of these results, the relation $\dot M_{\rm baryons}\gg\dot M_{\rm SIDM}$ holds true, as expected.

Another point to highlight is the lifetime of the SF soliton cloud. For $\lambda\sim10^{-19}$, $m\sim10^{-5}$ eV, and SMBH $\sim10^6M_\odot$, we obtained an accretion rate of $\dot M_{\rm SIDM}\sim10^{-10}M_\odot~{\rm yr}^{-1}$ and a scalar soliton mass of $\sim 10^{10} M_{\odot}$. We can see that the mass lost by the soliton through the accretion process is negligible, and as reported in \cite{Brax:2019npi,Feng:2021qkj}, with this accretion rate, the soliton has a much longer lifetime than the current age of the Universe.

\section{DISCUSSION AND CONCLUSION}
\label{sec:IV}

We are currently in an exciting era of high-precision observations of BH at the horizon scale. These observations, taking place in the strong field regime, have the potential to provide unprecedented insights not only into the properties of BH themselves but also into the environments in which they reside. A key premise underlying this scenario is that DM could play a crucial role on these observations, leaving distinctive signatures that may help reveal its elusive nature. But, can we truly extract valuable information about DM properties from BH observations? Moreover, what insights can we expect from upcoming observations? An encouraging finding suggests that BHs and DM can form stable, long-lived configurations, which have significant implications for astrophysical timing observations.

In this paper, we investigated the impact of BH charge on the accretion process of self-interacting SF onto a RN BH. Building upon previous studies, we have extended previous analysis to encompass a more general class of BHs, although we have opted to use Schwarzschild coordinates. To ensure the validity of our results, we derived all analytical expressions from the ground up in this new coordinate system and spacetime geometry.
Motivated by recent results from the EHT collaboration that allowed the existence of charged BHs within the current uncertainties, we have primarily focused on exploring the effects of BH charge within the range $q \in [0,9]$ to assess its influence on the energy flux. Through numerical analysis, we found that as $q$ increases, the energy flux is reduced by up to 20\% for the maximum allowable charge. Moreover, the charge also affects critical values $k_{c}$ and $x_\star$, resulting in their respective increase and decrease. These results can be inferred from Table \ref{resumen}.



Another interesting result we found is that the critical values $k_c$ and $x_\star$, which define the critical flow $F_c$, lie on the same curve, unlike the solutions $k_2$ and $k_1$, which are independent of the value of $q$. This behavior is illustrated in Fig.  \ref{Relacion k vs x}. However, the value of $q$ does shift the position of the couple ($x_\star$,$k_c$) along the critical curve. In this regard, the behavior of the flow is transonic, meaning that it selects a single physical solution that smoothly and continuously connects both branches of velocities (low and high). Furthermore, we observed that $x_\star$ is always smaller than $r_{\rm {isco}}$, implying that the maximum flow occurs on marginally bound orbits.

The most significant change in the SF profile occurs near the horizon when varying the charge. This is primarily due to the position of the horizon, which also affects the stability of the critical point in the accretion flow. We expected a relatively constant density around the horizon, as opposed to an abrupt increase due to the chosen singularity coordinate. As reported in \cite{Brax:2019npi} and evidenced here, for large distances, $\langle \rho_\phi \rangle \propto r^{-1}$. However, there is a slight correction arising from the charge, resulting in a small increase of the SF profile.

Additionally, we derived a new constraint on the model parameters based on the mass accretion rate of M87$^{\star}$ obtained from the EHT observations. Making the reasonable assumption that $\dot M_{\text{baryons}} \gg \dot M_{\text{DM}}$ and considering the observations of M87$^{\star}$, we placed constraints on the ($m$, $\lambda_4$) parameter space, which falls within the exclusion zone (see Fig. \ref{constrain lambda y masa} along with Eq. (\ref{constrain ecuacion})). It should be noted that this constraint is currently weak, but it may be improved with future measurements. There is a notable difference in the percentage of $\dot M_{\text{SIDM}}$ between an uncharged BH and one with a significant charge ($q = 0.9$), amounting to approximately 20\%. This implies that the accretion onto a charged BH is reduced due to the smaller size of the external horizon. Consequently, the region where self-interacting SF infalls is also diminished. Nevertheless, the value of $\dot M_{\text{SIDM}}$ remains on the order of $\sim 10^{-10} M_{\odot}$ yr$^{-1}$, which is subdominant compared to $\dot M_{\text{baryons}}$. Finally, we made a rough estimate suggesting that the soliton cloud formed inside galaxies remains stable under this level of accretion, indicating a significantly longer lifetime than the current age of the universe. Consequently, a more significant effect of the soliton clouds around BHs could potentially be observed in future high-precision observations.

In conclusion, the ongoing and forthcoming BH observations offer a unique opportunity to explore the role of DM in the dynamics of BHs and their surroundings. These observations, combined with theoretical advancements, will provide crucial insights into the properties of DM and its interactions with BHs. In this sense, our study has provided valuable insights into the impact of BH charge on the accretion process of the self-interacting SF onto RN BHs. We expect our findings to open up new avenues for future research in this exciting field of study.

\section{ACKNOWLEDGMENTS}
Y. R is supported by Beca Doctorado Convenio Marco de la Universidad de Santiago de Chile (USACH) from 2019-2020 and by Beca Doctorado Nacional año 2021 Folio  No. 21211644 de la Agencia Nacional de Investigacion y Desarrollo de Chile (ANID). G. G acknowledges financial support from Agencia Nacional de Investigaci\'on y Desarrollo (ANID), Chile, through the FONDECYT postdoctoral Grant No. 3210417. G. G also acknowledges Patrick Valageas for a valuable guidance at the beginning of this project and useful comments on this manuscript. N.C acknowledges the support of Universidad de Santiago de Chile (USACH), through Proyecto DICYT N° 042131CM, Vicerrectoría de Investigación, Desarrollo e Innovación.

\appendix*
\label{Apendice}

\section{SCALAR FREE CASE $\lambda_{4}\to 0$}
In this appendix, we show that the noninteracting case can be derived consistently from Eq. (\ref{Perfil densidad Schwarzschild}). As $\lambda_{4}$ approaches zero, $k$ must also strictly do so. It is important to observe that the latter implies that we are in the free case, rather than in the nonrelativistic scenario. This inference also holds in the large mass limit  Eq. (\ref{k2 muy chico}). Using this relation and taking the limit $k\to 0$, we arrive at the following conditions: ${\rm cn} \to {\rm cos} $, ${\bf K}(0) = \pi/2$, $\alpha = 0$ and $\phi_{0} = \sqrt{2 \rho}/m$. These conditions allow us to derive the density profile for the scalar free case
\begin{equation}
    \rho_{\phi} = \rho \left[ \frac{(2-f)}{f} \sin^2(mt-s) + \cos^2(mt-s)\right].
    \label{eq:rho SF libre}
\end{equation}
A similar result for the free case was obtained in \cite{Brax:2019npi} using a different coordinate system.

\bibliography{sample}

\end{document}